\documentclass[12pt]{article}
\usepackage{epsfig}
\usepackage{graphicx}
\usepackage{amsmath}

\begin{document}

 \newcommand{\beq}{\begin{equation}}
\newcommand{\eeq}{\end{equation}}
\newcommand{\bea}{\begin{eqnarray}}
\newcommand{\eea}{\end{eqnarray}}
\newcommand{\beqn}{\begin{eqnarray}}
\newcommand{\eeqn}{\end{eqnarray}}
\newcommand{\beas}{\begin{eqnarray*}}
\newcommand{\eeas}{\end{eqnarray*}}
\newcommand{\defi}{\stackrel{\rm def}{=}}
\newcommand{\non}{\nonumber}
\newcommand{\bquo}{\begin{quote}}
\newcommand{\enqu}{\end{quote}}
\newcommand{\qt}{\tilde q}
\newcommand{\m}{\tilde m}
\newcommand{\trho}{\tilde{\rho}}
\newcommand{\tn}{\tilde{n}}
\newcommand{\tN}{\tilde N}
\newcommand{\gsim}{\lower.7ex\hbox{$\;\stackrel{\textstyle>}{\sim}\;$}}
\newcommand{\lsim}{\lower.7ex\hbox{$\;\stackrel{\textstyle<}{\sim}\;$}}


\def\de{\partial}
\def\Tr{ \hbox{\rm Tr}}
\def\const{\hbox {\rm const.}}
\def\o{\over}
\def\im{\hbox{\rm Im}}
\def\re{\hbox{\rm Re}}
\def\bra{\langle}\def\ket{\rangle}
\def\Arg{\hbox {\rm Arg}}
\def\Re{\hbox {\rm Re}}
\def\Im{\hbox {\rm Im}}
\def\diag{\hbox{\rm diag}}


\def\QATOPD#1#2#3#4{{#3 \atopwithdelims#1#2 #4}}
\def\stackunder#1#2{\mathrel{\mathop{#2}\limits_{#1}}}
\def\stackreb#1#2{\mathrel{\mathop{#2}\limits_{#1}}}
\def\Tr{{\rm Tr}}
\def\res{{\rm res}}
\def\Bf#1{\mbox{\boldmath $#1$}}
\def\balpha{{\Bf\alpha}}
\def\bbeta{{\Bf\beta}}
\def\bgamma{{\Bf\gamma}}
\def\bnu{{\Bf\nu}}
\def\bmu{{\Bf\mu}}
\def\bphi{{\Bf\phi}}
\def\bPhi{{\Bf\Phi}}
\def\bomega{{\Bf\omega}}
\def\blambda{{\Bf\lambda}}
\def\brho{{\Bf\rho}}
\def\bsigma{{\bfit\sigma}}
\def\bxi{{\Bf\xi}}
\def\bbeta{{\Bf\eta}}
\def\d{\partial}
\def\der#1#2{\frac{\d{#1}}{\d{#2}}}
\def\Im{{\rm Im}}
\def\Re{{\rm Re}}
\def\rank{{\rm rank}}
\def\diag{{\rm diag}}
\def\2{{1\over 2}}
\def\ntwo{${\mathcal N}=2\;$}
\def\nfour{${\mathcal N}=4\;$}
\def\none{${\mathcal N}=1\;$}
\def\ntwot{${\mathcal N}=(2,2)\;$}
\def\ntwoo{${\mathcal N}=(0,2)\;$}
\def\x{\stackrel{\otimes}{,}}

\def\ba{\beq\new\begin{array}{c}}
\def\ea{\end{array}\eeq}
\def\be{\ba}
\def\ee{\ea}
\def\stackreb#1#2{\mathrel{\mathop{#2}\limits_{#1}}}

\def\Tr{{\rm Tr}}
\newcommand{\cpn}{CP$(N-1)\;$}
\newcommand{\wcpn}{wCP$_{N,\tilde{N}}(N_f-1)\;$}
\newcommand{\wcpd}{wCP$_{\tilde{N},N}(N_f-1)\;$}
\newcommand{\vp}{\varphi}
\newcommand{\pt}{\partial}
\newcommand{\ve}{\varepsilon}
\renewcommand{\theequation}{\thesection.\arabic{equation}}

\setcounter{footnote}0

\vfill

\begin{titlepage}

\begin{flushright}
FTPI-MINN-12/11, UMN-TH-3038/12\\
\end{flushright}

\vspace{1mm}

\begin{center}
{  \Large \bf  
\boldmath{$r$}-Duality and ``Instead-of-Confinement'' 
\\[0.5mm] 
Mechanism in  \boldmath{\none} 
Supersymmetric QCD
}

\vspace{1mm}

 {\large \bf    M.~Shifman$^{\,a}$ and \bf A.~Yung$^{\,\,a,b}$}
\end {center}

\begin{center}

$^a${\it  William I. Fine Theoretical Physics Institute,
University of Minnesota,
Minneapolis, MN 55455, USA}\\
$^{b}${\it Petersburg Nuclear Physics Institute, Gatchina, St. Petersburg
188300, Russia
}
\end{center}


\begin{center}
{\large\bf Abstract}
\end{center}

We consider \ntwo SQCD with the U$(N)$ gauge group and $N_f$ flavors 
($N_f>N$) perturbed by an \ntwo breaking deformation --
a small mass term $\mu$ for the adjoint matter. We study $r$-vacua, 
with the constraint $\frac{2}{3}N_f < r \le N$. At large values of the
parameter $\xi\sim\mu m$ ($m$ is a typical value of the quark masses) $r$ quark flavors condense, by construction.
The effective low-energy theory with the gauge group
U$(r)\times$U(1)$^{N-r}$ is at weak coupling. Upon reducing $\xi$ the original theory 
undergoes   a crossover transition from weak to strong coupling. 
 
As the original theory becomes strongly coupled, at low energies it is
 described by a weakly coupled infrared-free {\em dual}
 theory with the gauge group  U$(N_f-r)\times$U(1)$^{N-N_f+r}$ and  $N_f$ light dyon flavors.
These dyons condense triggering formation of non-Abelian strings which 
still confine monopoles, rather than quarks, contrary
to naive duality arguments. ``Instead-of-confinement'' mechanism for quarks and gauge bosons of the original theory
takes place:
screened quarks and gauge bosons of the original theory
decay, on curves of the marginal stability (CMS),  into confined monopole-antimonopole pairs   
that form stringy mesons. 

Next, we increase the deformation parameter $\mu$ thus
decoupling the adjoint fields. Then our theory flows to \none SQCD. 
The gauge group of the dual theory becomes U$(N_f-r)$. We show that
the dual theory is weakly coupled if we are sufficiently
close to the Argyres--Douglas point. The ``instead-of-confinement'' mechanism for quarks and gauge bosons 
survives in the limit of large $\mu$. It determines low-energy non-Abelian dynamics in 
the $r$-vacua of \none SQCD.

\vspace{2cm}

\end{titlepage}

 \newpage

\tableofcontents

\newpage

\section {Introduction }
\label{intro}
\setcounter{equation}{0}

The mechanism of confinement based on the  monopole condensation  \cite{mandelstam} was shown 
to work \cite{SW1,SW2} in the
monopole vacua of \ntwo supersymmetric QCD. This confinement {\em per se} is essentially 
Abelian \cite{DS,HSZ,Strassler,VY}. Non-Abelian gauge group is broken down to an Abelian subgroup 
by condensation of the adjoint scalars at a high scale, with the subsequent monopole condensation at 
a much lower scale,
 in a low-energy Abelian theory.
Simultaneously,
formation of confining flux tubes (strings) occurs.

In \none supersymmetric QCD there are no adjoint scalars.  One may hope that, starting 
from \ntwo QCD and decoupling the adjoint scalars, one can arrive at a 
regime with non-Abelian confinement.

Motivated by this idea we
recently  found \cite{SYdual,SYN1dual} a novel   non-Abelian duality 
 in the quark vacuum of \ntwo supersymmetric QCD with the U$(N)$
gauge group and $N_f$ flavors of fundamental matter (quarks), 
where $N <N_f<\frac32 N$. 
The theory was perturbed \cite{SYN1dual}
by a mass term $\mu$ for the adjoint matter. At small $\mu$ the deformation 
superpotential  reduces to the
Fayet--Iliopoulos (FI) \cite{FI} $F$-term with the effective FI parameter $\xi$
determined by $\xi\sim \mu m$, where $m$ presents a typical
scale of the quark masses.  In \cite{SYdual,SYN1dual} we  focused exclusively 
on the so-called $r=N$  vacuum in which $r=N$ quarks condense, thus completely
Higgsing the U$(N)$ gauge group.
A global color-flavor locked symmetry survives in the limit of equal quark mass terms.

At large $\xi$ this theory is at weak coupling and 
supports non-Abelian flux tubes (strings)
 \cite{HT1,ABEKY,SYmon,HT2} (for reviews see also \cite{Trev,Jrev,SYrev,Trev2}). It is the
formation of these strings that ensures confinement of monopoles. Monopoles manifest themselves 
 as junctions of two  different
strings. If $\xi \gg \Lambda_{{\mathcal N}=2}^2$, the problem can be treated quasiclassically
(here $\Lambda_{{\mathcal N}=2}$ is the scale of \ntwo SQCD).

Now, what happens if the value of $\xi$ decreases?
Upon reducing  the $\xi$ parameter, the theory
undergoes  a crossover transition \cite{SYdual,SYtorkink,SYcross} in a strongly 
coupled regime.  Needless to say, quasiclassical description
in terms of the original theory fails. 

At small $\xi$, dynamics  can be described
in terms of a weakly coupled {\em dual} \ntwo SQCD,
 with the U$(N_f-N)\times$U(1)$^{2N-N_f}$ gauge group and $N_f$ flavors of
light {\em dyons}.\footnote{This is in perfect agreement with the
results obtained in \cite{APS} where the SU$(N_f-N)$ dual  gauge group  was identified
at the root of the baryonic Higgs branch in the  SU($N$) gauge theory with massless (s)quarks.}
This structure is similar to
 Seiberg's duality in \none theories \cite{Sdual,IS}
where emergence of the dual SU$(N_f-N)$ group was first observed.

The dual theory supports non-Abelian strings due to condensation of light dyons in  much the same way
as the string formation in the original theory is due to condensation of squarks. Importantly, the
strings  of the dual theory confine monopoles, 
rather than quarks \cite{SYdual}.
This is due to the fact that the  light dyons that  condense in the dual theory  
carry weight-like chromoelectric charges (in addition to chromomagnetic charges). In other words, they carry
the quark charges. The strings formed through condensation of these dyons
can confine only the states with the root-like magnetic charges, i.e. 
the monopoles  \cite{SYdual}.  Thus, 
our \ntwo non-Abelian duality is {\em not}  electromagnetic. 

 Then, there  is no confinement of the chromoelectric charges; on the contrary, they
are {\em Higgs-screened}.

\begin{figure}
\epsfxsize=6cm
\centerline{\epsfbox{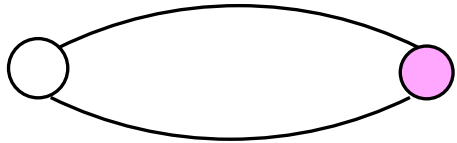}}
\caption{\small Meson formed by a monopole-antimonopole pair connected by two strings.
Open and closed circles denote the monopole and antimonopole, respectively.}
\label{figmeson}
\end{figure}

At strong coupling  where the dual description is applicable, the 
quarks and gauge bosons of the original theory are in what we call  
``instead-of-confinement'' phase. Namely,
the quarks and
gauge bosons decay into mono\-pole-antimonopole 
pairs on the curves of marginal stability (CMS) \cite{SYdual,SYtorkink}.
The (anti)monopoles forming the pair are confined. In other words, the original quarks and gauge bosons 
evolve in the 
strong coupling domain of small $\xi$  to become stringy mesons with two constituents 
being connected by two strings as shown in  Fig.~\ref{figmeson}.  
These mesons are expected to lie on Regge trajectories. 

Moreover, deep in the non-Abelian quantum regime the confined mono\-poles were demonstrated \cite{SYtorkink} 
to belong to the {\em fundamental representation} of the global  (color-flavor locked) group. 
Therefore, the  monopole-antimonopole
mesons can be both, in the adjoint and singlet representation of this  group. 
This  pattern  seems to be similar to what we have in actuality. The
 role of the ``constituent quarks'' inside the mesons is played by the monopoles.

The above referred to small values of the deformation parameter $|\mu |$. 
Next, we increased   its value, thus  decoupling the
adjoint fields and sending the original theory  to the limit of \none SQCD. 
At large $\mu$ the dual theory was demonstrated \cite{SYN1dual} to be   weakly coupled and 
infrared  (IR) free, with the 
 U$(N_f-N)$  gauge group  and $N_f$ light dyons $D^{lA}$, (here $l=1, ...,N_f-N$ is the color index 
in the dual gauge group, while $A=1,...,N_f$ is the flavor index). Our 
proof is valid  provided that the dyon condensate 
$\sim \xi \sim \mu m$ is small enough which, in turn,  requires the quark masses to be
 small in the large $\mu$ limit.  
Non-Abelian strings (albeit this time non-BPS saturated)
still confine monopoles while the
 quark and gauge bosons of  original \none SQCD  are presented by stringy mesons built from the
monopole-antimonopoles pairs connected by two non-Abelian strings, see Fig.~\ref{figmeson}.

``Instead-of-confinement'' mechanism is still at work.

\vspace{2mm}

In this paper we make a next step by exploring  other vacua of the $\mu$-deformed \ntwo theory, with 
 the number of condensed quarks $r$  smaller than $N$. Namely, we focus on the interval
\beq
\frac{2}{3}N_f < r \le N\,.
\label{rrange}
\eeq
The difference between these $r$ vacua from that with $r=N$ is that for  $r<N$ a U(1)  factor 
of the U$(N)$ gauge group always remains unbroken \cite{Cachazo2} and therefore
residual  long-range forces are present. 
The theory is not fully Higgsed.
Still we will show that the low-energy physics is rather similar to that of the  $r=N$ case. 

Strategically we follow the route similar to the analysis of
\cite{SYN1dual}. First we study non-Abelian duality at small $|\mu |$, not far from the \ntwo limit, 
and then increase $|\mu |$ sending the theory to  \none SQCD.
At large $\xi$ the low-energy physics is determined by a
weakly coupled  U$(r)\times$U(1)$^{N-r}$ gauge theory broken by the condensation of 
squarks down to U(1).

Upon reducing $\xi$ the theory goes through a crossover transition to 
strong coupling. At small $\xi$
the low-energy physics can be described by a dual weakly coupled IR free theory. 
The gauge group of the dual theory is 
\beq
U(\nu)\times U(1)^{N-\nu}, \qquad
\nu=\left\{
\begin{array}{cc}
r, & r\le \frac{N_f}{2}\\[2mm]
N_f-r, & r > \frac{N_f}{2} \\
\end{array}
\right. .
\label{nu}
\eeq
Given the constraint (\ref{rrange}) we focus on the case $\nu=N_f-r$. We will refer to this 
non-Abelian duality as ``$r$-duality.'' Only if $r=N$  our $r$-duality reduces to Seiberg-like duality
which we had studied in \cite{SYdual,SYN1dual}.

Note, that the presence of the SU$(\nu)\times $U(1)$^{N_f-\nu}$ gauge groups at 
the roots of the non-baryonic branches in massless ($\xi=0$)  \ntwo SU$(N)$ SQCD was first 
recognized  in \cite{APS}. Also, the relation between
$r$ and $\nu$ given in Eq.~(\ref{nu}) was  noted in \cite{Ookouchi,BolKonMar}, 
where it was interpreted as a correspondence
between ``classical and quantum $r$-vacua.'' Our interpretation is different:
we interpret it as a dual description emerging 
upon reducing $\xi$ below the crossover transition line.

Light matter of the dual theory is represented by $N_f$ flavors of dyons charged with respect to the gauge group
(\ref{nu}). We calculate their electric and magnetic charges and show that they are,
in fact,  quark-like states
with weight-like electric and root-like magnetic charges. Upon condensation of these dyons 
non-Abelian string are formed. We show explicitly that these strings confine 
monopoles, rather than quarks, in much the same way as in the $r=N$ vacuum. 

The distinction between the $r<N$ and $r=N$ vacua is that one  $Z_N$ string (let us say, the $N$-th,
there are $N$ 
$Z_N$ strings altogether)  is always absent
in the  $r<N$ vacua. The associated flux of the unbroken U(1) gauge factor is not confined. Instead, it
  is spread in accordance with the Coulomb law.
As a result, non-Abelian strings become metastable in the
$r<N$ vacua: they can be broken by monopole-antimonopole
pair creation, with  monopoles being junctions of  one of the first $r$ $Z_N$ strings and the would-be 
$N$-th string (which is in fact absent). At large quark masses these monopoles are heavy and strings are almost stable.

Next, we will
increase $\mu$ thus decoupling the adjoint matter,
 together with the U(1) factors of the dual gauge group (\ref{nu}) and singlet dyons. 
 The dual theory then reduces to a gauge theory with the gauge group 
\beq
U(\nu)\times U(1)^{{\rm unbr}}
\label{ggruplmu}
\eeq
and  $N_f$ non-Abelian quark-like dyons. Here U(1)$^{{\rm unbr}}$ denotes  the unbroken U(1) gauge factor.
Dyons are neutral with respect to U(1)$^{{\rm unbr}}$.
We integrate out heavy fields and present a superpotential for the
light dyons.
We show that this theory stays at week coupling as we increase $|\mu |$ provided that
we stay  close enough to the Argyres--Douglas (AD) point
\cite{AD} in the quark mass parameter space. Formation of 
the non-Abelian strings and monopole confinement ensue.

Our main results can be  summarized as follows. 

\begin{figure}
\epsfxsize=8cm
\centerline{\epsfbox{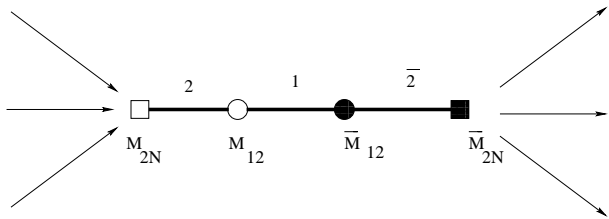}}
\caption{\small Example of the dipole meson formed as result of breaking of 2-nd string by pair creation of monopole
$M_{2N}$ (shown by boxes) interpolating between 2-nd string and would-be $N$-th string, which is absent.
Arrows denote unconfined flux.
Circles denote monopoles $M_{KK'}$, $K,K'=1,...,\nu$. 
Open and closed circles/boxes denote  monopoles and antimonopoles, respectively.}
\label{figdipole}
\end{figure}

We found that strongly coupled low-energy dynamics of \none supersym\-metric SQCD 
in the $r$-vacua in the range (\ref{rrange}) are {\em not} 
what one might naively expect  from electromagnetic
duality. The dual gauge group is U$(\nu )$ (where $\nu=N_f-r$) with $N_f$ flavors of light quark-like dyons.
Their condensation leads to formation of non-Abelian strings which still confine monopoles, 
rather than quarks.
The quarks and gauge bosons 
of the original theory
are in the  ``instead-of-confinement'' phase: upon crossing CMS from weak to strong
coupling
they decay 
into confined monopole-antimonopole pairs that 
form stringy mesons. For $r<N$ the
strings in the stringy mesons depicted in Fig.~\ref{figmeson} can be broken by a pair creation of particular monopoles which 
interpolate between the $K$-th string ($K=1,...,\nu$) and 
the  would-be $N$-th string, which is in fact absent. An example of the meson emerging in this way
is shown in Fig.~\ref{figdipole}. 

The endpoints emit fluxes of the unbroken U(1) gauge field. This makes this meson a dipole-like configuration. 
Note that the non-Abelian fluxes of the SU$(\nu)$ gauge group are always trapped and squeezed
in the non-Abelian strings.
Long-range forces are associated only with the unbroken U(1) gauge factor.
Monopoles inside the dipole meson cannot annihilate
if the overall flavor representation of the meson is nontrivial, say, the meson is in adjoint.

In a forthcoming publication \cite{SYvsS} we will compare the $r$-duality with {\em Seiberg's }
duality \cite{Sdual,IS}. 

To this end we will
consider a generalization \cite{CKM} of Seiberg's duality  to $r$ vacua (originally Seiberg's duality was formulated for 
the monopole $r=0$ vacua).
 In the $r=N$ vacuum our dual gauge group U$(\nu=N_f-r)$ coincides with Seiberg's dual group U$(N_f-N)$. 
Moreover, in this case Seiberg's dual superpotential has a classical vacuum. 

 We will show that,
upon integrating out heavy mesonic $M$-fields, this superpotential coincides with our dual superpotential obtained in \cite{SYN1dual}, while Seiberg's ``dual quarks'' in fact reduce to our quark-like dyons $D^{lA}$.

At the same time, in the window $\frac{2}{3}N_f <r<N$ vacua 
our $r$ duality does not match Seiberg's duality. 
Our dual theory has the
U$(\nu)$ gauge group instead of U$(\tN)$ and a
different superpotential for light matter. 
Our dual theory does have a supersymmetric classical vacuum
 and, in a certain regime (with small $\xi$), stays at weak coupling. Thus, it is appropriate to speak of
  triality.

For the $r$ vacua in the range  $\frac{2}{3}N_f <r<N$  Seiberg's dual superpotential has no supersymmetric classical vacua if the quark  mass
terms are nonvanishing. Integrating out Seiberg's ``dual quarks'' 
one obtains a continuation of the Afleck--Dine--Seiberg  superpotential 
\cite{ADS} to $N_f>N$.
This superpotential correctly reproduces the quark and 
gaugino condensates and gives the correct number of the quantum vacua \cite{CKM,GivKut}. 

We interpret this as follows \cite{SYvsS}.
In the $r$ vacua in the range $\frac{2}{3}N_f < r < N$ the generalized Seiberg dual theory 
is in fact in the strong coupling regime and therefore is not useful in
 describing low-energy physics in its entirety. 
However, it does describe the chiral sector in the sense of the
Veneziano--Yankielowicz effective superpotential \cite{Ven} (which is not a genuine
low-energy superpotential). Namely, 
 chiral condensates are correctly reproduced. 
 The spectrum of excitations is not.
 
Low-energy physics in the $r$ vacua  is described (in the range $\frac{2}{3}N_f < r < N$)  
by weakly coupled $r$-dual theory with the dual gauge group U$(\nu=N_f-r)$ rather than U$(N_f-N)$.

We also show in \cite{SYvsS} that classical supersymmetric vacua of Seiberg's 
dual theory detected in \cite{CKM,GivKut} correspond to  
smaller $r$, namely to $r<(N_f-N)$. In this range Seiberg's dual theory is at weak coupling 
and hence describes low-energy physics in full. This range, however, is beyond the scope of the present paper. 
 
In this paper we only consider the $r$-vacua in the range (\ref{rrange}). The detailed study of 
the $r$-vacua with $r\le \frac23 N_f$
is left for future work. Still, we make a few qualitative comments about these vacua. Our picture suggests 
that we have a conformal window
in the $r$-vacua in the range 
\beq
\frac13 N_f\le r\le \frac23 N_f\,.
\label{cwindow}
\eeq
This means that even if we take  \none SQCD with  $N<N_f<\frac32 N$, the $r$-vacua in the range (\ref{cwindow})
are described by a conformal theory in the IR.

If $r<\frac13 N_f$ then Eq.~(\ref{nu}) gives $\nu=r$; therefore, there is no crossover transition upon reducing $\xi$.
The dual theory has the same gauge group U$(r)$ as the original one. This suggests that in the dual theory we have 
a regular Higgs phase for quarks, and ``instead-of-confinement'' mechanism does not work. 
Quarks and gauge bosons at strong coupling are just Higgs-screened, rather than transformed into 
 stringy mesons of the type shown in Fig.~\ref{figmeson} or Fig.~\ref{figdipole}.

A  problem for future studies is extrapolating our construction of
$r$ duality to $r\le \frac{2}{3}N_f$ and comparing it 
in this range with Seiberg's duality, in particular, of importance is 
the range $r<(N_f-N)$ where the Seiberg's dual theory is at weak coupling.

The paper is organized as follows. In Sec.~\ref{bulk} we describe our basic theory, $\mu$-deformed 
\ntwo SQCD.\footnote{
For a detailed review of this model see \cite{SYrev}.} 
In Sec.~\ref{r=N}, as a preparation for original explorations,  we summarize what is known
about the non-Abelian duality and ``instead-of-confinement'' mechanism in the  $r=N$ vacuum. Then, 
in Sec.~\ref{rduality}, we proceed to the $r$-duality. We consider the Seiberg--Witten curve and derive Eq.~(\ref{nu}).
Section~\ref{r=N-1largexi} is devoted to a thorough  study of the $r=N-1$ vacuum. In this particular
example we describe in detail
the low-energy theory at large $\xi$ and in the small-$\mu$ limit. The passage to still smaller $r$ becomes
qualitatively clear.
In Sec.~\ref{rdual} we reduce $\xi$ and calculate the  light dyon charges in the dual theory. Monopole confinement
is demonstrated.
We  present  the action of the dual theory and use exact Seiberg--Witten curves to calculate 
the vacuum expectation values (VEVs) of the 
dyon fields. In Sec.~\ref{rdualitylargemu} we increase the value of the 
deformation parameter $\mu$, decouple the adjoint matter 
and derive effective superpotential for light non-Abelian dyons. 
Section~8 summarizes our conclusions. In Appendices A--D we present calculational details of our analysis.

\section {Basic Model: \boldmath{$\mu$}-Deformed \ntwo SQCD}
\label{bulk}
\setcounter{equation}{0}

The gauge symmetry of our basic model is 
U($N$)=SU$(N)\times$U(1). In the absence
of  deformation the model under consideration is \ntwo  SQCD
 with $N_f$ massive quark hypermultiplets. 
 We assume that
$N_f>N$ but $N_f<\frac32 N$. 
The latter inequality ensures  that the dual theory can be infrared free. 

In addition, we will introduce the mass term $\mu$ 
for the adjoint matter breaking \ntwo supersymmetry down to \none. 

The field content is as follows. The \ntwo vector multiplet
consists of the  U(1)
gauge field $A_{\mu}$ and the SU$(N)$  gauge field $A^a_{\mu}$,
where $a=1,..., N^2-1$, and their Weyl fermion superpartners plus
complex scalar fields $a$, and $a^a$ and their Weyl superpartners, respectively.
The $N_f$ quark multiplets of  the U$(N)$ theory consist
of   the complex scalar fields
$q^{kA}$ and $\tilde{q}_{Ak}$ (squarks) and
their   fermion superpartners --- all in the fundamental representation of 
the SU$(N)$ gauge group.
Here $k=1,..., N$ is the color index
while $A$ is the flavor index, $A=1,..., N_f$. We will treat $q^{kA}$ and $\tilde{q}_{Ak}$
as rectangular matrices with $N$ rows and $N_f$ columns. 

Let us first discuss the undeformed  \ntwo theory.
 The  superpotential has the form
 \beq
{\mathcal W}_{{\mathcal N}=2} = \sqrt{2}\,\sum_{A=1}^{N_f}
\left( \frac{1}{ 2}\,\tilde q_A {\mathcal A}
q^A +  \tilde q_A {\mathcal A}^a\,T^a  q^A + m_A\,\tilde q_A q^A\right)\,,
\label{superpot}
\eeq
where ${\mathcal A}$ and ${\mathcal A}^a$ are  chiral superfields, the ${\mathcal N}=2$
superpartners of the gauge bosons of  U(1) and SU($N$), respectively.

Next, we add a mass term for the adjoint fields which breaks \ntwo
supersymmetry down to \none,
\beq
{\mathcal W}_{{\rm br}}=\sqrt{\frac{N}{2}}\,\frac{\mu_0}{2} {\mathcal A}^2
+  \frac{\mu}{2}({\mathcal A}^a)^2,
\label{msuperpotbr}
\eeq
where $\mu_0$ and $\mu$ is are mass parameters for the chiral
superfields in \ntwo gauge supermultiplets,
U(1) and SU($N$), respectively.\footnote{Without loss of generality one can assume them to be real.} In this paper 
we will  consider the single-trace perturbation which amounts to choosing $\mu_0$ such, that 
the parameter
\beq
\gamma = 1-\sqrt{\frac2N}\frac{\mu_0}{\mu}=0.
\label{gamma}
\eeq
Clearly, the mass term (\ref{msuperpotbr}) splits the 
\ntwo supermultiplets, breaking
\ntwo supersymmetry down to \none. 

The bosonic part of the action of our basic 
theory has the form  (for details see \cite{SYrev})
\beqn
S&=&\int d^4x \left[\frac1{4g^2_2}
\left(F^{a}_{\mu\nu}\right)^2 +
\frac1{4g^2_1}\left(F_{\mu\nu}\right)^2
+
\frac1{g^2_2}\left|D_{\mu}a^a\right|^2 +\frac1{g^2_1}
\left|\partial_{\mu}a\right|^2 \right.
\nonumber\\[4mm]
&+&\left. \left|\nabla_{\mu}
q^{A}\right|^2 + \left|\nabla_{\mu} \bar{\tilde{q}}^{A}\right|^2
+V(q^A,\tilde{q}_A,a^a,a)\right]\,.
\label{model}
\eeqn
Here $D_{\mu}$ is the covariant derivative in the adjoint representation
of  SU$(N)$, while
\beq
\nabla_\mu=\partial_\mu -\frac{i}{2}\; A_{\mu}
-i A^{a}_{\mu}\, T^a
\label{defnab}
\eeq
acts in the fundamental representation.
We suppress the color  SU($N$)  indices of the matter fields. The normalization of the 
 SU($N$) generators  $T^a$ is as follows
$$
{\rm Tr}\, (T^a T^b)=\mbox{$\frac{1}{2}$}\, \delta^{ab}\,.
$$
The coupling constants $g_1$ and $g_2$
correspond to the U(1)  and  SU$(N)$  sectors, respectively.
With our conventions, the U(1) charges of the fundamental matter fields
are $\pm1/2$, see Eq.~(\ref{defnab}).

The scalar potential $V(q^A,\tilde{q}_A,a^a,a)$ in the action (\ref{model})
is the  sum of  $D$ and  $F$  terms,
\beqn
V(q^A,\tilde{q}_A,a^a,a) &=&
 \frac{g^2_2}{2}
\left( \frac{1}{g^2_2}\,  f^{abc} \bar a^b a^c
 +
 \bar{q}_A\,T^a q^A -
\tilde{q}_A T^a\,\bar{\tilde{q}}^A\right)^2
\nonumber\\[3mm]
&+& \frac{g^2_1}{8}
\left(\bar{q}_A q^A - \tilde{q}_A \bar{\tilde{q}}^A \right)^2
\nonumber\\[3mm]
&+& 2g^2_2\left| \tilde{q}_A T^a q^A 
+\frac{1}{\sqrt{2}}\,\,\frac{\pt{\mathcal W}_{{\rm br}}}{\pt a^a}\right|^2+
\frac{g^2_1}{2}\left| \tilde{q}_A q^A +\sqrt{2}\,\,\frac{\pt{\mathcal W}_{{\rm br}}}{\pt a} \right|^2
\nonumber\\[3mm]
&+&\frac12\sum_{A=1}^{N_f} \left\{ \left|(a+\sqrt{2}m_A +2T^a a^a)q^A
\right|^2\right.
\nonumber\\[3mm]
&+&\left.
\left|(a+\sqrt{2}m_A +2T^a a^a)\bar{\tilde{q}}^A
\right|^2 \right\}\,.
\label{pot}
\eeqn
Here  $f^{abc}$ denote the structure constants of the SU$(N)$ group,
$m_A$ is the mass term for the $A$-th flavor,
 and 
the sum over the repeated flavor indices $A$ is implied.

The  vacua of the theory (\ref{model}) are determined by the zeros of 
the potential (\ref{pot}). In general, the theory has a number of the so called $r$-vacua, in which 
(quasiclassically) $r$ squarks condense. Later we will show that this quasiclassical analysis is
 valid if we require the parameter $\xi\sim \mu m$ to be large, with $m$ being a typical scale of 
 the quark masses.
The overall range of variation of $r$ is  $r=0, ..., N$. Say, the
$r=0$ vacua (there are $N$ such vacua)  are always at strong
coupling. These are in fact the monopole vacua of \cite{SW1,SW2}. 

\section {Duality in the \boldmath{$r=N$} vacuum}
\label{r=N}
\setcounter{equation}{0}

In this section we will briefly review non-Abelin duality in the $r=N$ vacua established
 in \cite{SYdual,SYtorkink,SYN1dual}.
These vacua have  the maximal possible number of condensed quarks, $r=N$. Moreover,
  the gauge group U$(N)$ is completely 
Higgsed in these vacua, and, as a result, 
they support non-Abelian strings \cite{HT1,ABEKY,SYmon,HT2}. The occurrence of 
these strings ensures 
confinement of the monopoles in these vacua. 

First, we  will assume that $\mu$ is small, much smaller than
the quark masses
\beq
\mu\ll | m_A |, \qquad A=1, ..., N_f\,.
\label{smallmu}
\eeq

\subsection{Vacuum structure at large \boldmath{$\xi$}}

Now we  assume  that our theory is at weak coupling,  so that we can  
analyze it quasiclassically. 
With generic values of the quark masses we have 
\beq
C_{N_f}^{N}= \frac{N_f!}{N!(N_f-N)!}
\label{numva}
\eeq
 isolated $r$-vacua in which $r=N$ quarks (out of $N_f$) develop
vacuum expectation values  (VEVs).
Consider, say, the vacuum in which the first $N$ flavors develop VEVs, to be denoted as (1, 2 ..., $N$).
In this vacuum  the
adjoint fields  develop  
VEVs too, namely,
\beq
\left\langle \Phi\right\rangle = - \frac1{\sqrt{2}}
 \left(
\begin{array}{ccc}
m_1 & \ldots & 0 \\
\ldots & \ldots & \ldots\\
0 & \ldots & m_N\\
\end{array}
\right),
\label{avev}
\eeq
where
\beq
\Phi \equiv \frac12\, a + T^a\, a^a \,.
\label{Phi}
\eeq
For generic values of the quark masses, the  SU$(N)$ subgroup of the gauge 
group is
broken down to U(1)$^{N-1}$. However, in the {\em special limit} of equal masses,
\beq
m_1=m_2=...=m_{N_f}\,,
\label{equalmasses}
\eeq
the  adjoint field VEVs do not break the SU$(N)\times$U(1) gauge group.
In this limit the theory acquires  a global flavor SU$(N_f)$ symmetry.

With all quark masses equal (and limiting ourselves to the leading order in $\mu$),
 the mass term for the adjoint matter (\ref{msuperpotbr})
reduces to the Fayet--Iliopoulos $F$-term of the U(1) factor of the SU$(N)\times$U(1) gauge group,
  which does {\em not} break \ntwo supersymmetry \cite{HSZ,VY}. 
Higher orders in the parameter $\mu$
  break \ntwo supersymmetry by splitting all \ntwo multiplets.
  
If the quark masses are unequal the U($N$) gauge group is broken  down to U(1)$^{N}$
by the adjoint field VEVs (\ref{avev}). 

Using (\ref{msuperpotbr}) and (\ref{avev}) it is not difficult to obtain the quark field VEVs
 from Eq.~(\ref{pot}).   Up to  a gauge rotation they can be written as \cite{SYfstr}
\beqn
\langle q^{kA}\rangle &=& \langle\bar{\tilde{q}}^{kA}\rangle=\frac1{\sqrt{2}}\,
\left(
\begin{array}{cccccc}
\sqrt{\xi_1} & \ldots & 0 & 0 & \ldots & 0\\
\ldots & \ldots & \ldots  & \ldots & \ldots & \ldots\\
0 & \ldots & \sqrt{\xi_N} & 0 & \ldots & 0\\
\end{array}
\right),
\nonumber\\[4mm]
k&=&1,..., N\,,\qquad A=1,...,N_f\, ,
\label{qvev}
\eeqn
where we present the quark fields as  matrices in the color ($k$) and flavor ($A$) indices.
The Fayet--Iliopoulos $F$-term parameters for each U(1) gauge factor are given (in the quasiclassical
approximation) by the following expressions:
\beq
\xi_P \approx 2\;\mu m_P,
\qquad P=1,...,N.
\label{xis}
\eeq

While the adjoint VEVs do not break the SU$(N)\times$U(1) gauge group in the limit
(\ref{equalmasses}), the quark condensate (\ref{qvev}) does result in  the spontaneous
breaking of both gauge and flavor symmetries.
A diagonal global SU$(N)$ combining the gauge SU$(N)$ and an
SU$(N)$ subgroup of the flavor SU$(N_f)$
group survives, provided that the quark masses are equal. 
This is color-flavor locking. Below we will refer to this diagonal
global symmetry as to $ {\rm SU}(N)_{C+F}$.

Thus, the pattern  of the
color and flavor symmetry breaking
is as follows: 
\beq
{\rm U}(N)_{\rm gauge}\times {\rm SU}(N_f)_{\rm flavor}\to  
{\rm SU}(N)_{C+F}\times  {\rm SU}(N_f-N)_F\times {\rm U}(1)\,.
\label{c+f}
\eeq
Here SU$(N)_{C+F}$ is a global unbroken color-flavor rotation, which involves the
first $N$ flavors, while the SU$(N_f-N)_F$ factor stands for the flavor rotation of the 
$(N_f-N)$ quarks.
As we will see shortly, the global symmetry of the dual theory is, of course, 
the same, albeit the physical origin is different.
The presence of the global SU$(N)_{C+F}$ group is instrumental for
formation of the non-Abelian strings \cite{HT1,ABEKY,SYmon,HT2,SYfstr}.
Tensions of $N$ elementary strings are determined \cite{SYfstr} by 
the parameters $\xi_P$, see (\ref{xis}),
\beq
T_P=2\pi\xi_P\,.
\label{ten}
\eeq

Since the global (flavor) SU$(N_f)$ group is broken by the quark VEVs anyway, it will be helpful for 
our purposes
to consider
the following mass splitting:
\beq
m_P=m_{P'}, \qquad m_K=m_{K'}, \qquad m_P-m_K=\Delta m
\label{masssplit}
\eeq
where 
\beq
 P, P'=1, ... , N\,\,\,\, {\rm and}   \,\,\,\, K, K'=N+1, ... , N_f\,.
 \label{pppp}
\eeq
This mass splitting respects the global
group (\ref{c+f}) in the $(1,2,...,N)$ vacuum. Moreover, this vacuum  becomes  isolated.
No Higgs branches develop. We will often focus on this limit below in this section.

Now, let us briefly discuss the perturbative excitation spectrum. 
Since
both U(1) and SU($N$) gauge groups are broken by the squark condensation, all
gauge bosons become massive.

To the leading order in $\mu$, \ntwo supersymmetry is not broken.  In fact, with nonvanishing $\xi_P$'s (see Eq.~(\ref{xis})), both the quarks and adjoint scalars  
combine  with the gauge bosons to form long \ntwo supermultiplets \cite{VY},  for a review see \cite{SYrev}.
In the limit (\ref{masssplit}) $\xi_P\equiv\xi\,,$  and all states come in 
representations of the unbroken global
 group (\ref{c+f}), namely, in the singlet and adjoint representations
of SU$(N)_{C+F}$,
\beq
(1,\, 1), \quad (N^2-1,\, 1),
\label{onep}
\eeq
 and in the bifundamental representations
\beq
 \quad (\bar{N},\, N_f-N), \quad
(N,\, \bar{N_f}-\bar{N})\,.
\label{twop}
\eeq
We mark representations in (\ref{onep}) and (\ref{twop})  with respect to two 
non-Abelian factors in (\ref{c+f}). The singlet and adjoint fields are (i) the gauge bosons, and
(ii) the first $N$ flavors of the squarks $q^{kP}$ ($P=1,...,N$), together with their fermion superpartners.
The bifundamental fields are the quarks $q^{kK}$ with $K=N+1,...,N_f$.
These quarks transform in the two-index representations of the global
group (\ref{c+f}) due to the color-flavor locking. Singlet and adjoint fields have masses of order $g\sqrt{\xi}$,
while masses of bifundamental fields are equal to $\Delta m$.

The above quasiclassical analysis is valid if the theory is at weak coupling. This is the case if
the quark VEVs are sufficiently large so that the  gauge coupling constant is frozen at a large scale.
From (\ref{qvev}) we see that 
the quark condensates are of the order of
$\sqrt{\mu m}$ (see also \cite{SW1,SW2,APS,CKM}). The weak 
coupling condition reduces to
\beq
\sqrt{\mu m}\gg\Lambda_{{\mathcal N}=2}\,,
\label{weakcoup}
\eeq
where $\Lambda_{{\mathcal N}=2}$ is the scale of the \ntwo  theory, and we assume that all quark masses are of the same order $m_A\sim m$. 
In particular,  the condition (\ref{weakcoup}), combined with the condition (\ref{smallmu}) of smallness
of $\mu$,   implies that the average quark mass $m$ is very large. 

\subsection{Dual theory}
\label{vacjumpN}

Now we will relax the condition (\ref{weakcoup}) and pass
to the strong coupling domain at 
\beq
|\sqrt{\xi_P}|\ll \Lambda_{{\mathcal N}=2}\,, \qquad | m_{A}|\ll \Lambda_{{\mathcal N}=2}\,,
\label{strcoup}
\eeq
still keeping $\mu$ small.
 
In \cite{SYdual,SYN1dual}
it was shown that the theory (\ref{model}) in the $r=N$ vacuum undergoes a
crossover transition as the value of $\xi$ decreases.
The  domain (\ref{strcoup}) 
can be described in terms of weakly coupled (infrared free) dual theory
with with the {\em gauge group}
\beq
{\rm U}(N_f-N)\times {\rm U}(1)^{2N-N_f}\,,
\label{dualgaugegroup}
\eeq
 and $N_f$ flavors of light {\em dyons}.\footnote{ Previously the SU$(N_f-N)$
 gauge group  was identified  \cite{APS} as dual  on the Coulomb   branch 
at the root of the baryonic Higgs branch in the \ntwo supersymmetric  SU($N$) Yang--Mills 
theory with massless quarks.}

\vspace{2mm}

Light dyons $D^{lA}$ 
($l=1, ... ,(N_f-N)$ and $A=1, ... , N_f$) are in 
the fundamental representation of the gauge group
SU$(N_f-N)$ and are charged under the Abelian factors indicated in Eq.~(\ref{dualgaugegroup}).
 In addition, there are  $(2N-N_f)$ 
light dyons $D^J$ ($J=(N_f-N+1), ... , N$), neutral under 
the SU$(N_f-N)$ group, but charged under the
U(1) factors.

The dyon condensates are as follows:
\beqn
\!\!\!\!
\langle D^{lA}\rangle \!\!\! \!& =& \langle \bar{\tilde{D}}^{lA}\rangle \! =
\frac1{\sqrt{2}}\,\left(
\begin{array}{cccccc}
0 & \ldots & 0 & \sqrt{\xi_{1}} & \ldots & 0\\
\ldots & \ldots & \ldots  & \ldots & \ldots & \ldots\\
0 & \ldots & 0 & 0 & \ldots & \sqrt{\xi_{(N_f-N)}}\\
\end{array}
\right)\!,
\label{Dvev}
\\[4mm]
\langle D^{J}\rangle &=& \langle\bar{\tilde{D}}^{J}\rangle=\sqrt{\frac{\xi_J}{2}}, 
\qquad J=(N_f-N +1), ... , N\,.
\label{adiiz}
\eeqn
The most important feature apparent in (\ref{Dvev}), as compared to the squark VEVs  in the 
original theory (\ref{qvev}),  is a ``vacuum leap'' \cite{SYdual},
\beq
(1, ... ,\, N)_{\sqrt{\xi}\gg \Lambda_{{\mathcal N}=2}} \to 
(N+1, ... , \,N_f,\,\,(N_f-N+1), ... ,\, N)_{\sqrt{\xi}\ll \Lambda_{{\mathcal N}=2}}\,.
\label{jump}
\eeq
In other words, if we pick up the vacuum with nonvanishing VEVs of the  first $N$ quark flavors
in the original theory at large $\xi$, Eq.~(\ref{model}),  and then reduce $\xi$ below 
$\Lambda_{{\mathcal N}=2}$, 
the system goes through a crossover transition and ends up in the vacuum of the {\em dual} theory with
the nonvanishing VEVs of $(N_f-N) $ last dyons (plus VEVs of $(2N-N_f)$ dyons that are SU$ (N_f-N)$ singlets).

The Fayet--Iliopoulos parameters $\xi_P$  in (\ref{Dvev}), (\ref{adiiz}) are determined by the 
quantum version of the classical expressions
(\ref{xis}) \cite{SYfstr}.   
Defining
\beq
u_k= \left\langle {\rm Tr}\left(\frac12\, a + T^a\, a^a\right)^k\right\rangle, \qquad k=1, ..., N\,,
\label{u}
\eeq
we perform a quantum generalization in  the two relevant terms in the third line 
of the potential in (\ref{pot}),
\beq
\frac{\pt{\mathcal W}_{{\rm br}}}{\pt a^a}\to\mu \,\frac{\pt u_2}{\pt a^a}\,,
\qquad \quad\frac{\pt{\mathcal W}_{{\rm br}}}{\pt a}\to\mu \,\frac{\pt u_2}{\pt a}\,.
\label{dwda}
\eeq
From this we  obtain 
\cite{SYfstr}
\beq
\xi_P=-2\sqrt{2}\,\mu\,E_P\,,
\label{qxis}
\eeq
where $E_P$ ($P=1, ..., N$) are the diagonal elements of the $N\times N$ matrix
\beq
E=\frac1{N}\,\frac{\pt u_2}{\pt a}+T^{\tilde{a}}\,\frac{\pt u_2}{\pt a^{\tilde{a}}}\,.
\label{E}
\eeq
Here $T^{\tilde{a}}$ are the Cartan generators of the SU$(N)$ gauge group (the subscript $\tilde{a}$ runs over
 $\tilde{a}=1,...,(N-1)$).

The parameters $E_P$ are expressible  in  terms of the roots of the Seiberg--Witten curve. Namely,
in the given $r=N$ vacuum  they are  \cite{SYfstr}
\beq
E_P=e_P, \qquad P=1, ... , N\, ,
\label{ErN}
\eeq
where $e_P$ are the double roots of the Seiberg--Witten curve \cite{APS}, 
\beq
y^2= \prod_{P=1}^{N} (x-\phi_P)^2 -
4\left(\frac{\Lambda_{{\mathcal N}=2}}{\sqrt{2}}\right)^{2N-N_f}\, \,\,\prod_{A=1}^{N_f} \left(x+\frac{m_A}{\sqrt{2}}\right),
\label{curve}
\eeq
while $\phi_P$ are gauge invariant parameters on the Coulomb branch.

In the $r=N$ vacuum the curve (\ref{curve}) has $N$ double roots and reduces to
\beq
y^2= \prod_{P=1}^{N} (x-e_P)^2,
\label{rNcurve}
\eeq
where quasiclassically (at large masses) $e_P$'s are given  by the
mass parameters, $\sqrt{2}e_P\approx -m_P$ ($P=1, ... , N$).

Thus, the dyon condensates at small $\xi$ in the $r=N$ vacuum are determined by 
\beq
\xi_P=-2\sqrt{2}\,\mu\,e_P\,.
\label{xirN}
\eeq
We will see below that the expressions (\ref{qvev}), (\ref{Dvev}) and (\ref{qxis}) are quite general 
and valid also for the $r<N$ vacua, while the relation (\ref{ErN}) gets modified in the $r<N$ vacua.

As long as we keep $\xi_P$ and masses small enough (i.e. in the domain (\ref{strcoup}))
the coupling constants of the
infrared-free dual theory (frozen at the scale of the dyon VEVs) are small:
the dual theory is at weak coupling.

At small masses, in the region  (\ref{strcoup}), the double roots of the Seiberg--Witten
 curve are  
\beq
\sqrt{2}e_I = -m_{I+N}, \qquad 
\sqrt{2}e_J = \Lambda_{{\mathcal N}=2}\,\exp{\left(\frac{2\pi i}{2N-N_f}J\right)}
\label{roots}
\eeq
for $2N-N_f>1$, where 
\beq
I=1, ... ,(N_f-N)\,\,\,\, {\rm  and} \,\,\,\, J=(N_f-N+1), ... , N\,.
\label{d1}
\eeq
 In particular, the $(N_f-N)$ first roots are determined by the masses of the 
 last $(N_f-N)$ quarks --- a reflection of the fact that the 
non-Abelian sector of the dual theory is not asymptotically free and is at weak coupling
in the domain (\ref{strcoup}). 

\subsection{``Instead-of-confinement'' mechanism}
\label{instoc}

Now, let us consider either the equal quark masses or the special choice (\ref{masssplit}).
Both, the gauge group and the global flavor SU($N_f$) group, are
broken in the vacuum. In the case of (\ref{masssplit}) the flavor SU($N_f$) group is explicitly
broken down to SU($N)\times$SU$(N_f-N)$ by masses.
 However, the color-flavor locked form apparent in (\ref{Dvev})
under the given mass choice guarantees that the diagonal
global SU($N_f-N)_{C+F}$ symmetry survives. More exactly, the  unbroken {\em global} group of the dual
theory is 
\beq
 {\rm SU}(N)_F\times  {\rm SU}(N_f-N)_{C+F}\times {\rm U}(1)\,.
\label{c+fd}
\eeq
The SU$(N_f-N)_{C+F}$ factor in (\ref{c+fd}) is a global unbroken color-flavor rotation, which involves the
last $(N_f-N)$ flavors, while the SU$(N)_F$ factor stands for the flavor rotation of the 
first $N$ dyons. 

Thus,  color-flavor locking takes place in the dual theory too. In much the same way as 
in the original theory, the presence of the global SU$(N_f-N)_{C+F}$ symmetry
is the  reason behind formation of the non-Abelian strings. Their tensions are still given by Eq.~(\ref{ten}),
where the parameters $\xi_P$ are determined by (\ref{xirN}) \cite{SYfstr,SYN1dual}.
 For generic quark masses the  global symmetry  (\ref{c+f}) is broken down to 
U(1)$^{N_f-1}$. 

In the equal mass limit, or given the special choice (\ref{masssplit}),
the global unbroken symmetry (\ref{c+fd}) of the dual theory at small
$\xi$ coincides with the global group (\ref{c+f}) which manifests itself in the
$r=N$ vacuum of the original theory at large
$\xi$.  

Note, however, that this global symmetry is realized in two very distinct ways in the dual pair at hand.
As was already mentioned, the quarks and U($N$) gauge bosons of the original theory at large $\xi$
come in the following representations 
 of the global group (\ref{c+f}):
 $$
 (1,1), \,\, (N^2-1,1), \,\, (\bar{N},(N_f-N)), \,\,{\rm and} \,\, (N,(\bar{N_f}-\bar{N}))\,.
 $$
At the same time,  the dyons and U($N_f-N$) gauge 
bosons of the dual theory form 
$$(1,1),\,\, (1,(N_f-N)^2-1),\,\, (N,(\bar{N_f}-\bar{N})), \,\, {\rm and}\,\,
(\bar{N},(N_f-N))$$ 
representations of (\ref{c+fd}). We see that the
adjoint representations of the $(C+F)$
subgroup are different in two theories. How can this happen?
     
The quarks and gauge bosons
which form the  adjoint $(N^2-1)$ representation  
of SU($N$) at large $\xi$ and the dyons and gauge bosons which form the  adjoint $((N_f-N)^2-1)$ representation  of SU($N_f-N$) at small $\xi$ are, in fact, {\em distinct} states.
The $(N^2-1)$  adjoints of SU($N$) become heavy 
and decouple as we pass from large to small $\xi$ 
 along the line $\xi\sim \Lambda_{{\mathcal N}=2}$. Moreover, some 
composite $((N_f-N)^2-1)$ adjoints  of SU($N_f-N$), which are 
heavy  and invisible in the low-energy description at large $\xi$ become light 
at small $\xi$ and form the $D^{lK}$ dyons
 ($K=N+1, ... , N_f$) and gauge bosons of U$(N_f-N)$. The phenomenon of 
 the level crossing
 takes place. Although this crossover is smooth in the full theory,
from the standpoint of the low-energy description the passage from  large to small $\xi$  means a dramatic change: the low-energy theories in these domains are 
completely
different; in particular, the degrees of freedom in these theories are different.

This logic leads us to the following conclusion \cite{SYdual}. In addition to light dyons and gauge bosons 
included in  the low-energy theory at small $\xi$ we must have
heavy  fields  which form the adjoint representation $(N^2-1,1)$ of the 
global symmetry (\ref{c+fd}). These are screened  quarks 
and gauge bosons from the large-$\xi$ domain.

As has been already noted in Sec.~\ref{intro},
at small $\xi$ they decay into the monopole-antimonopole 
pairs on the curves of marginal stability (CMS).\footnote{An explanatory remark regarding
our terminology is in order. Strictly speaking,
such pairs  can be  formed by monopole-antidyons and
 dyon-antidyons as well, the dyons carrying root-like electric charges. 
 In this paper we refer to all such states collectively as to
``monopoles." This is to avoid confusion with dyons which appear in Eq.~(\ref{Dvev}). The
latter dyons carry weight-like electric charges and, roughly speaking, behave as
quarks, see \cite{SYdual} for further details.}
This is in accordance with the results obtained in \cite{SW1,SW2,BF}
for \ntwo SU(2) gauge theories,  on the Coulomb branch at vanishing $\xi$.
For the theory at hand this picture was established  in \cite{SYtorkink}.
The general rule is that the only states that exist at strong coupling inside CMS are those which can become massless on the Coulomb branch
\cite{SW1,SW2,BF}. For our theory these are light dyons shown in Eq.~(\ref{Dvev}),
gauge bosons of the dual gauge group and monopoles.
 
 At small nonvanishing values of $\xi$ the
monopoles and antimonopoles produced in the decay process of the adjoint $(N^2-1,1)$ states
cannot escape from
each other and fly off to asymptotically large separations 
because they are confined. Therefore, the (screened) quarks or gauge bosons 
evolve into stringy mesons  in the  strong coupling domain of small  $\xi$ 
-- the  monopole-antimonopole
pairs connected  by two strings \cite{SYdual,SYN1dual}, as shown in  Fig.~\ref{figmeson}. 
This is what we call ``instead-of-confinement'' mechanism for quarks and gauge bosons.

\subsection{\boldmath{$r=N$} Duality at large $\mu$}

From Eqs.~(\ref{Dvev}), (\ref{qxis}) and (\ref{roots}) we see that the VEVs of the non-Abelian
dyons $D^{lA}$ are determined by $\sqrt{\mu m}$ and are much smaller 
than the VEVs of the Abelian dyons $D^{J}$ in the domain
(\ref{strcoup}).  The latter are of the order of 
$\sqrt{\mu \Lambda_{{\mathcal N}=2}}$. 
This circumstance is most crucial for us. It  allows us to  increase $\mu$
and decouple the adjoint fields without spoiling the weak coupling condition in the dual theory \cite{SYN1dual}.

Now we assume that 
\beq
|\mu| \gg |m_A |, \qquad A=1, ... , N_f\,.
\label{mularge}
\eeq
The VEVs of the Abelian dyons become large at large $\mu$. This makes U(1) gauge fields of the dual group 
(\ref{dualgaugegroup}) heavy. Decoupling these gauge factors, together with 
the adjoint matter and 
the Abelian dyons themselves, we obtain the low-energy theory with the
\beq
U(N_f-N)
\label{dualgglmurN}
\eeq
gauge fields and the non-Abelian dyons $D^{lA}$ ($l=1, ... , N_f-N$, $\,A=1, ... , N_f$).
For the single-trace perturbation (\ref{msuperpotbr}) with $\gamma=0$ the superpotential 
for $D^{lA}$ has the form \cite{SYN1dual}
\beq
{\mathcal W} = -\frac1{2\mu}\,
(\tilde{D}_A D^B)(\tilde{D}_B D^A)  
+m_A\,(\tilde{D}_A D^A)\,,
\label{superpotd}
\eeq
where the color indices are contracted inside each parentheses.

The minimization of this superpotential leads to the dyon VEVs,  
\beq
\langle D^{lA}\rangle \! \! = \langle \bar{\tilde{D}}^{lA}\rangle =
\!\!
\frac1{\sqrt{2}}\,\left(
\begin{array}{cccccc}
0 & \ldots & 0 & \sqrt{\xi_{1}} & \ldots & 0\\
\ldots & \ldots & \ldots  & \ldots & \ldots & \ldots\\
0 & \ldots & 0 & 0 & \ldots & \sqrt{\xi_{(N_f-N)}}\\
\end{array}
\right),
\label{DvevN1}
\eeq
where those $\xi$'s that enter Eq. (\ref{DvevN1}) (cf. Eq. (\ref{Dvev}))  are of the order of $\mu m$, see (\ref{roots}).
Other $\xi$'s (see Eq. (\ref{adiiz})) become irrelevant, since all U(1) gauge fields  become heavy at large $\mu$
and decouple.

Below the scale $\mu$ our theory becomes dual to \none SQCD with the scale
\beq
\tilde{\Lambda}^{3N-2N_f}= \frac{\Lambda_{{\mathcal N}=2}^{2N-N_f}}{\mu^{N_f-N}}\,.
\label{tildeL}
\eeq
The only condition we impose to keep this infrared-free theory in the weak coupling 
regime is 
\beq
|\sqrt{\mu m}| \ll \tilde{\Lambda}\,.
\label{wcdual}
\eeq
This means that at large $\mu$ we must keep the quark masses sufficiently small. 

We would like to stress that if VEV's of dyons were all of order  of $\sqrt{\mu \Lambda_{{\mathcal N}=2}}$,
it would not be  possible to  decouple 
the adjoint matter keeping the dual theory at weak coupling. Once we 
increased $\mu$ above the scale $\sqrt{\mu \Lambda_{{\mathcal N}=2}}$, we would get that these VEVs 
are much larger than $\tilde{\Lambda}$, which breaks the weak coupling 
condition in the dual theory. 
Thus, the non-Abelian structure present in the dual theory 
is the most important element of the continuation to large $\mu$.

To summarize, at large $\mu$ and small $\xi$ the original \none SQCD in the $r=N$ vacuum goes 
through a crossover transition
at strong coupling. In the domain (\ref{wcdual}) it is described by the 
weakly coupled infrared-free  dual theory, U$(N_f-N)$ SQCD, with   
$N_f$ light dyon flavors. Condensation of the light dyons
$D^{lA}$ in this theory triggers formation of the non-Abelian strings and confinement of monopoles.
For quarks and gauge bosons
 of the original \none SQCD we have an ``instead-of-confinement'' phase: they decay into the
 monopole-antimonopole pairs on CMS and 
form stringy mesons shown in Fig.~\ref{figmeson}.

\section {\boldmath{$r$}-Duality}
\label{rduality}
\setcounter{equation}{0}

Now we are finally ready to turn to the main topic of this paper -- the study of the $r<N$ vacua.
First we consider the small-$\mu$ domain in which the theory is  close to the \ntwo limit.
Our task  is to analyze the transition  from  large  to small $\xi$. In
much the same way as in \cite{SYdual} we  will   do this in two steps.
First, we will assume the quark mass differences to be large. In this domain the theory stays at
weak coupling, and we can safely decrease  the value of the  parameter $\xi$. Next,
we will use the exact Seiberg--Witten solution of the theory on the Coulomb branch \cite{SW1,SW2}
(i.e. at   $\xi\to 0$) to perform the passage from the domain of the
large quark mass differences to the domain of the small quark mass differences. 

With  large mass differences, the quark sector of the theory 
in the $r$-vacuum is at weak coupling and can be analyzed semiclassically.
The number of the $r$-vacua with $r<N$ in our theory is \cite{CKM}
\beq
(N-r)\,C_{N_f}^{r}= (N-r)\,\frac{N_f!}{r!(N_f-r)!}\, ,
\label{nurvac}
\eeq
i.e. is equal to the number of choices one can pick up $r$ quarks which develop VEVs (out of $N_f$ quarks) times the
Witten index (the number of vacua) in the classically unbroken SU$(N-r)$ pure gauge theory. 

Below we consider a particular vacuum where the {\em first} $r$ quarks develop 
VEVs (cf. Sec.~\ref{r=N}), 
to be labeled by
$(1, ... ,\, r)$. Quasiclassically at large mass differences the VEVs of 
the adjoint scalars are given by
\beq
\left\langle {\rm diag}\left(\frac12\, a + T^a\, a^a\right)\right\rangle \approx - \frac1{\sqrt{2}}
\left[m_1,...,m_r,0,...,0 
\right],
\label{avevr}
\eeq
where the first $r$  diagonal elements are  proportional to the quark masses, while the last $(N-r)$ entries   classically
vanish.
In quantum theory they become of order of $\Lambda_{{\mathcal N}=2}$.

Now we have to identify this vacuum in terms of the Seiberg--Witten curve. 
In our theory (\ref{model}) it has the form \cite{APS}
\beq
y^2= \prod_{k=1}^{N} (x-\phi_k)^2 -
4\left(\frac{\Lambda_{{\mathcal N}=2}}{\sqrt{2}}\right)^{2N-N_f}\, \,\,\prod_{A=1}^{N_f} \left(x+\frac{m_A}{\sqrt{2}}\right),
\label{curve1}
\eeq
where $\phi_k$ are gauge invariant parameters on the Coulomb branch. Semiclassically,
\beq
{\rm diag}\left(\frac12\, a + T^a\, a^a\right) \approx 
\left[\phi_1,...,\phi_N\right].
\eeq
Therefore, in  the ($1, ... ,\, r$) quark vacuum we have
\beq
\phi_P \approx -\frac{m_P}{\sqrt{2}},\quad P=1, ... ,\, r\,, \qquad 
\phi_P \sim \Lambda_{{\mathcal N}=2},\quad P=r+1, ... ,\, N\,
\label{classphi}
\eeq
in the large $m_A$ limit, see (\ref{avevr}). 

To identify the $r<N$ vacuum in terms of the curve (\ref{curve1}) it is necessary to find
such values of $\phi_P$ which would ensure  the curve to have $N-1$ double roots. 
$r$ parameters $\phi_P$'s are determined by the quark masses in the semiclassical limit, 
see (\ref{classphi}).
$N-1$ double roots are associated with $r$ condensed quarks and $N-r-1$ condensed monopoles.
 Altogether, $N-1$ condensed states.
 
In contrast, in the $r=N$ vacuum we have the maximal possible number of the condensed states (quarks),  
namely, $N$ in U$(N)$ theory. This difference is related to the presence of the unbroken U(1) gauge group
in the $r<N$ vacua \cite{Cachazo2}. In the classically unbroken (after quark condensation) U$(N-r)$ gauge group
$N-r-1$ monopoles condense at the quantum level, thus breaking the non-Abelian SU$(N-r)$ subgroup. One U(1)
factor remains unbroken because the monopoles do not interact with it.

Now we pass to the limit of the equal quark masses (\ref{equalmasses}) and address the following question.
What is the maximal number of $\phi$'s which are determined by the quark masses exactly, without
$\Lambda_{{\mathcal N}=2}$ corrections? Let us denote this number by $\nu$. Let us rewrite the curve
(\ref{curve1}) as
\beqn
y^2
&=&
  \left(x+\frac{m}{\sqrt{2}}\right)^{2\nu} 
\nonumber\\[3mm]
&\times&
\left\{\prod_{k=\nu+1}^{N} (x-\phi_k)^2-
4\left(\frac{\Lambda_{{\mathcal N}=2}}{\sqrt{2}}\right)^{2N-N_f}\,  
\left(x+\frac{m}{\sqrt{2}}\right)^{N_f-2\nu}\right\},
\label{redcurve}
\eeqn
where the first $\nu$ $\phi$'s are given by
\beq
\phi_P = -\frac{m}{\sqrt{2}},\qquad P=1,...,\nu\,.
\label{firstphi}
\eeq
This curve has $\nu$ double roots located at
\beq
e_P =-\frac{m}{\sqrt{2}},\qquad P=1, ... ,\,\nu\,.
\label{firstroots}
\eeq

Now, the reduced curve in the curly brackets has $(N-\nu)$ colors and $(N_f-2\nu)$ flavors. If the maximal number
of quarks (all of them) condense in this reduced theory, the rank of the classically unbroken gauge group would be
$(N-\nu)-(N_f-2\nu)$. This number should be equal to the rank of the classically unbroken group in the $r$-vacuum
of the full theory. This gives
\beq
(N-\nu)-(N_f-2\nu) =N-r\,,
\eeq
which entails
\beq
\nu=N_f-r.
\eeq
Note, that the number of flavors in the reduced curve should be, of course, non-negative. This gives 
$N_f-2\nu\ge 0$ or $$r\ge N_f/2\,.$$ For smaller $r$ it is obvious that $\nu=r$. Thus, we arrive at
\beq
\nu=\left\{
\begin{array}{cc}
r, & r\le \frac{N_f}{2}\\[3mm]
N_f-r, & r > \frac{N_f}{2} \\
\end{array}
\right. \,.
\label{nu1}
\eeq

The main feature of the solution (\ref{firstphi}) is the absence of $O\left(\Lambda_{{\mathcal N}=2}\right)$ 
corrections to the first
$\nu$ $\phi$'s. This means that in the equal mass limit these $\nu$ $\phi$'s
become equal. This is a signal of restoration of the non-Abelian SU($\nu$) gauge group,
i.e. the gauge group of the dual theory at small $\xi$.

 Namely, the dual gauge group  in the equal mass limit becomes
\beq
{\rm U}(\nu)\times {\rm U}(1)^{N-\nu}.
\label{rdualgaugegroup}
\eeq
This is in perfect agreement with the
results obtained in \cite{APS,CKM} where  non-Abelian gauge groups  were identified
at the roots of the nonbaryonic Higgs branches in the SU($N$) gauge theory with 
the massless quarks.

The novel element of our analysis presented in this section is that
we started from the  non-Abelian $r$-vacuum at large $\xi$ and demonstrated that,
as we reduce $\xi$,
the theory in this vacuum undergoes crossover to a different non-Abelian regime,
with the dual low-energy gauge group (\ref{rdualgaugegroup}).
As was already mentioned, the physical 
reason for the emergence of the non-Abelian  gauge group is that the low-energy
effective theory  with the dual gauge group (\ref{rdualgaugegroup}) is infrared-free 
in the equal mass 
limit and  stays at weak coupling. Therefore, the classical analysis showing that the non-Abelian
gauge group is restored in the equal mass limit remains intact in
quantum theory.

As was already mentioned,  we interpret (\ref{nu1}) as a crossover transition 
with respect to the parameter $\xi$. If $r > N_f/2$ the
rank of the dual non-Abelian gauge subgroup SU$(\nu)$ at small $\xi$ is different from the rank of the original non-Abelian subgroup SU$(r)$. This difference imply a ``vacuum leap'' ( see Secs.~\ref{vacjumpN}
and~\ref{vjump}) and occurrence of ``instead-of-confinement'' mechanism. 

For $r < N_f/2$ there is no crossover.

\section {\boldmath{$r=N-1$} vacuum at large \boldmath{$\xi$}}
\label{r=N-1largexi}
\setcounter{equation}{0}

Our main example of the $r$ vacuum in this paper is 
\beq
r=N-1\,,
\eeq
 in the theory (\ref{model}).
We will use the same strategy as for the study of the $r=N$ vacuum: first assume  that $\mu$ is small
and the theory is close to the \ntwo limit, so we can use the exact Seiberg--Witten solution valid near the Coulomb branch.
We will study the crossover from the large-$\xi$ domain where the low-energy gauge group is 
\beq
{\rm U}(r=N-1)\times {\rm U}(1)^{\rm unbr}
\label{gaugegroupr}
\eeq
to the small-$\xi$ domain where the dual theory has the gauge group 
\beq
{\rm U}(\nu=N_f-N-1)\times {\rm U}(1)^{N-\nu-1}\times {\rm U}(1)^{\rm unbr}.
\label{gaugegrouprd}
\eeq
 At the last stage we will increase $\mu$ thus decoupling the adjoint matter.

Although in this paper we mostly consider the $r=(N-1)$ vacuum as a particular  
example of $r<N$ vacua in the theory
(\ref{model}),  we believe that our results are general and can be applied to all   $r$ vacua.

We also note, that while we keep $\mu$ small to ensure the proximity of the theory at hand to
the  \ntwo limit, we need a weaker condition
to have a crossover into strong coupling, namely $r>N_f/2$, see (\ref{nu1}). At the last stage,
in Sec.~\ref{rdualitylargemu}, we make $\mu$ large and assume that $r > \frac23\,N_f$ in order to keep the dual \none theory infrared free.

\subsection{Low-energy theory}

The low-energy theory in the $r=N-1$ vacuum at large $\xi$ is presented in Appendix A. It includes non-Abelian gauge fields
$A_{\mu}^{n}$ ($n=1, ... , r^2-1)$ as well as Abelian fields $A_{\mu}$ and $A_{\mu}^{N^2-1}$. 
The last one  is associated with the last Cartan generator of the SU$(N)$
group. These fields have scalar \ntwo superpartners $a^{n}$, $a$ and 
$a^{N^2-1}$. Light matter consists of quarks $q^{kA}$ ($k=1,...,r$). Note, that
all non-Abelian gauge fields from the sector SU$(N)$/SU$(r)$ are heavy and decouple in the large mass limit due to
 the structure of the adjoint VEV's (\ref{avevr}). Also $q^{NA}$ quarks  are heavy and not included in the 
 low-energy theory.

The potential (\ref{potr})  determines the vacuum structure in 
the $r=N-1$ vacuum. The adjoint VEV's have the form
\beq
\left\langle {\rm diag} \left( \Phi \right)\right\rangle \approx - \frac1{\sqrt{2}}
\left[m_1, ... , m_{N-1},0
\right],
\label{avevN-1}
\eeq
while the (s)quark VEV's are  
\beqn
\langle q^{kA}\rangle &=& \langle\bar{\tilde{q}}^{kA}\rangle=\frac1{\sqrt{2}}\,
\left(
\begin{array}{cccccc}
\sqrt{\xi_1} & \ldots & 0 & 0 & \ldots & 0\\
\ldots & \ldots & \ldots  & \ldots & \ldots & \ldots\\
0 & \ldots & \sqrt{\xi_{(N-1)}} & 0 & \ldots & 0\\
\end{array}
\right),
\nonumber\\[5mm]
k&=&1,..., (N-1)\,,\qquad A=1,...,N_f\, ,
\label{qvevr}
\eeqn
where now the first $(N-1)$  parameters $\xi$ are given quasiclassically by (\ref{xis}) while 
\beq
\xi_N =0\,.
\label{xiN}
\eeq
The last condition reflects the fact that the $N$-th quark is heavy and  develops no  VEV.

To see that this is the case we can use the general formula (\ref{qxis}) for $\xi$'s where the quasiclassical expression 
for the matrix $E$  
reduces to
\beq
 {\rm diag}\left(E \right) \approx
\left\langle {\rm diag} \left( \Phi \right)\right\rangle \approx - \frac1{\sqrt{2}}
\left[m_1,...,m_{N-1},0
\right].
\label{Ewcr}
\eeq

As is seen from Eq.~(\ref{defnablar}), the quarks interact with a particular linear combination
of the U(1) gauge fields $A_{\mu}$ and $A_{\mu}^{N^2-1}$, namely,
\beq
 A_{\mu} +\sqrt{\frac2{N(N-1)}}\; A_{\mu}^{N^2-1}.
\label{A0}
\eeq
The 	quark VEVs make this combination massive. The orthogonal combination
\beq
\sqrt{\frac2{N(N-1)}}\; A_{\mu} - A_{\mu}^{N^2-1}.
\label{unbroken}
\eeq
remains massless and corresponds to the unbroken U(1)$^{\rm unbr}$ gauge group.

In the equal mass limit the global flavor symmetry SU$(N_f)$ is broken in the $r$ vacuum
down to
\beq
{\rm SU}(r)_{C+F}\times  {\rm SU}(\nu=N_f-r)_F\times {\rm U}(1)\,.
\label{c+fr}
\eeq
Now SU$(r)_{C+F}$ is a global unbroken color-flavor rotation, which involves only the
first $r$ flavors, while the SU$(\nu=N_f-r)_F$ factor stands for the flavor rotation of the 
remainder of the  quark sector.

Since the global (flavor) SU$(N_f)$ group is broken by the quark VEVs anyway, 
it is useful  to consider the following mass splitting:
\beqn
&&
m_P=m_{P'}, \qquad m_K=m_{K'}, \qquad m_P-m_K=\Delta m,
\nonumber\\[3mm]
&&
 P, P'=1, ... , r\,\,\,\, {\rm and}   \,\,\,\, K, K'=r+1, ... , N_f\,.
 \label{masssplitr}
\eeqn
This mass splitting respects the global
group (\ref{c+fr}) in the $(1,2,...,r)$ vacuum. This vacuum  becomes  isolated.

In much the same way as in the $r=N$ vacuum, in the $r=N-1$ vacuum
   all states in the limit (\ref{masssplitr}) come in 
representations of the unbroken global
 group (\ref{c+fr}), namely, in the singlet and adjoint representations
of SU$(r)_{C+F}$,
\beq
(1,\, 1), \quad (r^2-1,\, 1),
\label{onepr}
\eeq
 and in the bifundamental representations
\beq
 \quad (\bar{r},\, \nu), \quad
(r,\, \bar{\nu})\,.
\label{twopr}
\eeq
We mark representations in (\ref{onep}) and (\ref{twop})  with respect to two 
non-Abelian factors in (\ref{c+fr}). The singlet and adjoint fields are the gauge bosons, and
 the first $r$ flavors of the quarks $q^{kP}$ ($P=1,...,r$). 
The bifundamental fields are the quarks $q^{kK}$ with $K=r+1,...,N_f$.
 Singlet and adjoint fields have masses of order $g\sqrt{\xi}$, where $\xi$ is the common value of 
the first $r$ parameters $\xi$ in the limit (\ref{masssplitr}),
while  the bifundamental field masses   are equal to $\Delta m$.

The above quasiclassical analysis applies provided that the theory is at weak coupling. The weak 
coupling condition is
\beq
|\sqrt{\xi}| \sim |\sqrt{\mu m}|\gg\Lambda_{{\mathcal N}=2}^{\rm LE}\,,
\label{weakcoupr}
\eeq
where $\Lambda_{{\mathcal N}=2}^{\rm LE}$ is the scale of the low-energy   theory (\ref{modelr}) determined by 
\beq
\Lambda_{{\mathcal N}=2}^{2N-N_f}=m^2 \,(\Lambda_{{\mathcal N}=2}^{\rm LE})^{2(N-1)-N_f}\,.
\label{LambdaN2LE}
\eeq

\subsection{Strings and confinement of monopoles at large \boldmath{$\xi$}}
\label{monconf}

As quarks develop VEVs in the $r=N-1$ vacuum the monopoles should be confined, in much  the same way
as they are in the $r=N$ vacuum. As was already mentioned, the distinction is that a single
 U(1) factor of the gauge group
remains unbroken; therefore the associated magnetic flux should be unconfined.
In this section we will determine the elementary string fluxes  in the classical limit 
at large $\xi$ to show that the elementary monopole
fluxes can be absorbed by two strings. Hence, 
the monopoles are indeed represented by the junctions of two different strings. The exceptions are
the  monopoles $M_{PN}$ ($P=1,...,r$)
interpolating between an $P$-th elementary string and the $N$-th would-be string (which is in fact absent).

To make our discussion simpler we will
consider here (and, often, below)  the theory with U$(N=4)$ gauge group and $N_f=5$ as an example,
\beq
N=4\, , \qquad N_f=5\,, \qquad r=3, \qquad \nu=2\,.
\label{example}
\eeq
In this case the low-energy theory (\ref{modelr}) has the
gauge group U(3)$\times$ U(1)$_{15}$, where U(1)$_{15}$ describes
 the gauge field $A_{\mu}^{N^2-1}$ with $N=4$. 

If the quark masses are unequal, the U(3) gauge group is broken down to U(1)$^3$ and 
the non-Abelian strings become $Z_{N=4}$ Abelian strings, see~\cite{SYrev} for more details.\footnote{One of these strings is absent in the $r=3$ vacuum.} Let us calculate their fluxes. Charges of three quarks $q^{kA}$, $k=1,2,3$ in (\ref{modelr}) can be written as
\beqn
&&\vec{n}_{q^1}=
\left(\frac12,0;\,\frac12,0;\,\frac1{2\sqrt{3}},0;\,\frac1{2\sqrt{6}},0\right), \nonumber\\[2mm]
&&\vec{n}_{q^2}=
\left(\frac12,0;\,-\frac12,0;\,\frac1{2\sqrt{3}},0;\,\frac1{2\sqrt{6}},0\right), \nonumber \\[2mm]
&&\vec{n}_{q^3}=
\left(\frac12,0;\,0,0;\,-\frac1{\sqrt{3}},0;\,\frac1{2\sqrt{6}},0\right),
\label{quarkcharges}
\eeqn
respectively, where we use the notation
\beq
\vec{n}=\left(n_e,n_m;\,n_e^3,n_m^3;\,n_e^8,n_m^8;\,n_e^{15},n_m^{15}\right),
\label{chargenotation}
\eeq
and
$n_e$ and $n_m$ denote electric and 
magnetic charges of a given state with respect to the U(1) gauge group, while $n_e^3$,
$n_m^3$, $n_e^8$, $n_e^8$ and $n_e^{15}$, $n_e^{15}$ stand for the electric and 
magnetic charges  with respect to the Cartan
generators of the SU(4) gauge group (broken down to U(1)$^3$ by quark mass differences).
In Appendix B  for convenience we present weights and roots of the SU(4) algebra. Quark charges correspond 
to the weights of this algebra. Note, that the 4-th quark is heavy and does not enter in the low-energy theory
(\ref{modelr}).

Consider one of the $Z_4$ strings which is formed due to the winding of the $q^{11}$ quark  at $r\to\infty$
(see \cite{SYrev,SYdual} for a more detailed discussion of the construction of the non-Abelian strings),
\beq
q^{11} \sim\sqrt{\frac{\xi_1}{2}}\,e^{i\alpha},
\qquad q^{22} \sim \sqrt{\frac{\xi_2}{2}}, \qquad q^{33}\sim \sqrt{\frac{\xi_3}{2}},
\label{qwind}
\eeq
see (\ref{qvevr}). Here $r$ and $\alpha$ are the polar coordinates in the plane $i=1,2$ orthogonal 
to the string axis. Note that in the $r=N=4$ vacuum there is one extra condition associated with the fourth quark
\cite{SYdual}. In the
$r=3$ vacuum this condition is absent.
Equations~(\ref{qwind})  imply the following behavior of the gauge potentials at $r\to\infty$:
\beqn
&& \frac12 A_i +\frac12 A_i^3 + \frac1{2\sqrt{3}} A_i^8 +\frac1{2\sqrt{6}} A_i^{15} \sim \pt_i \alpha\,,
\nonumber\\[2mm]
&& \frac12 A_i -\frac12 A_i^3 + \frac1{2\sqrt{3}} A_i^8 +\frac1{2\sqrt{6}} A_i^{15}\sim 0\,,
\nonumber\\[2mm]
&& \frac12 A_i  - \frac1{\sqrt{3}} A_i^8 +\frac1{2\sqrt{6}} A_i^{15}\sim 0\,,
\label{windings}
\eeqn
see the quark charges in (\ref{quarkcharges}). In the $r=3$ vacuum we have to supplement these conditions
with one extra condition which ensures that the combination (\ref{unbroken}) of 
the gauge potentials
$A_{\mu}$ and $A_{\mu}^{15}$, which has no interaction with quarks, is not excited, namely,
\beq
\frac1{\sqrt{6}}\,A_i - A_i^{15} \sim 0\,.
\label{ort=0}
\eeq
The solution to  equations (\ref{windings}) is
\beqn
&&  A_i  \sim \frac47 \,\pt_i \alpha\,, \qquad
A_i^3  \sim \pt_i \alpha\,,
\nonumber\\[2mm]
&&   A_i^8 \sim \frac1{\sqrt{3}}\,\pt_i \alpha\,, \qquad
 A_i^{15} \sim \frac{4}{7\sqrt{6}}\,\pt_i \alpha\,.
\label{gaugewindI}
\eeqn
It determines the string gauge  fluxes $\int dx_i A_i$, $\int dx_i A^3_i$, $\int dx_i A^8_i$ and 
$\int dx_i A^{15}_i$, respectively.
The integration above is performed over a large circle in the $(1,2)$ plane.
Let us call this string $S_1$.

Next, we define the string charges \cite{SYdual} as 
\beqn
&&
\int dx_i (A^D_i,\,A_i;\,A^{3D}_i,\,A^{3}_i;\,A^{8D}_i,\,A^{8}_i;\,A^{15D}_i,\,A^{15}_i)
\nonumber\\
&&
= 4\pi\,(-n_e,\,n_m;\,-n^3_e,\,n^3_m;-\,n^8_e,\,n^8_m;-\,n^{15}_e,\,n^{15}_m)\,.
\label{defstrch}
\eeqn
This definition ensures that the string has the same charge as a trial monopole which can be attached to
the string endpoint. In other words, the flux of the given  string is the flux of a trial 
monopole\,\footnote{This trial monopole does not necessarily exist in our
theory.  In the U$(N)$ theories the SU$(N)$ monopoles are rather string junctions, so they are 
attached to {\em two} strings, \cite{SYmon,SYdual}.} 
sitting on string's end, with
the charge defined by (\ref{defstrch}).

In particular, according to this definition, the charge of the string with the fluxes (\ref{gaugewindI})
is 
\beq
\vec{n}_{S_1}=\left(0,\,\frac27;\,0,\,\frac12;\,\,0,\,\frac1{2\sqrt{3}};\,\,0,\,\frac2{7\sqrt{6}}\right).
\label{S1}
\eeq
Since this string is formed through the quark condensation, it is  magnetic. 

There are two other  elementary strings $S_2$ and $S_3$  which arise due to winding of 
$q^{22}$ and $q^{33}$
quarks, respectively. Repeating the above procedure for these strings we get their charges,
\beqn
&&
\vec{n}_{S_2}= \left(0,\,\frac27;\,0,\,-\frac12;\,\,0,\,\frac1{2\sqrt{3}};\,\,0,\,\frac2{7\sqrt{6}}\right),
\nonumber\\[4mm]
&&
\vec{n}_{S_3}=\left(0,\,\frac27;\,0,\,0;\,\,0,\,-\frac1{\sqrt{3}};\,\,0,\,\frac2{7\sqrt{6}}\right).
\label{S23}
\eeqn
Note, that the fourth string $S_4$ of the U(4) gauge group is absent in the $r=3$ vacuum since the fourth quark is heavy,  have no VEV and, therefore, can  have no winding.

It is easy to check that each of the three elementary SU(4) monopoles associated with first three roots of 
the SU(4) algebra
(see Appendix B) is confined by two elementary
strings.  Consider, say, two elementary monopoles from  the
SU$(r=3)$ subgroup with  the charges $\vec{n}_{M_{12}}=(0,\,0;\,0,\,1;\,0,\,0;\,0,\,0)$
and $\vec{n}_{M_{23}}=(0,\,0;\,0,\,-\frac12;\,0,\,\frac{\sqrt{3}}{2};\,0,\,0)$. 
These charges can be written as
a difference of the charges of two elementary strings, namely, 
\beqn
&&
\vec{n}_{M_{12}}=(0,\,0;\,0,\,1;\,0,\,0;\,0,\,0)= \vec{n}_{S_1}-\vec{n}_{S_2}\,,
\nonumber\\[2mm]
&&
\vec{n}_{M_{23}}=(0,\,0;\,0,\,-\frac12;\,0,\,\frac{\sqrt{3}}{2};\,0,\,0)= \vec{n}_{S_2}-\vec{n}_{S_3}\, .
\label{conflargexi}
\eeqn
This means that each of these monopoles  (at large $\xi$) is in fact a junction of two strings, 
with one string having
the outgoing flux while
the other    incoming. The third $M_{13}$ monopole from the SU$(r=3)$ subgroup can 
be considered as a bound state of two elementary ones in (\ref{conflargexi}).

So far the monopole confinement in the $r=N-1$ vacuum looks quite similar to that in the
$r=N$ vacuum \cite{SYdual}.
The distinction becomes apparent once we consider the SU$(N=4)$ monopole which does not belong to 
the SU$(r=3)$ subgroup.
Let us consider the $M_{34}$ monopole with charges
\beq
\vec{n}_{M_{34}}=\left(0,\,0;\,0,\,0;\,0,\,-\frac1{\sqrt{3}};\,0,\,\sqrt{\frac23}\right).
\label{M34}
\eeq
In the $r=4$ vacuum this monopole is a junction of two strings $S_3$ and $S_4$. In the $r=3$ vacuum 
the $S_4$ string is absent. Let us calculate the unconfined  flux of the $S_3$ string with the monopole $M_{34}$ attached to its end. To this end consider the difference
\beq
\vec{n}_{\rm unconf} = \vec{n}_{S_3}-\vec{n}_{M_{34}}\,= \frac{2\sqrt{6}}{7}\,
\left(0,\,\frac1{\sqrt{6}};\,0,\,0;\,0,\,0;\,0,\,-1\right).
\label{unconfflux}
\eeq
We see that the $n^8_m$ charge is cancelled, and the resulting charge is a source of the U(1) gauge 
magnetic field corresponding to the following combination:
\beq
\frac1{\sqrt{6}}\,A_{\mu} - A_{\mu}^{15}, 
\label{unconf}
\eeq
This is exactly the field of the unbroken U(1)$^{\rm unbr}$ gauge group, see (\ref{unbroken}). 

Thus, the $S_3$
string  can terminate on the $M_{34}$ monopole  producing a magnetic source for the unbroken 
U(1)$^{\rm unbr}$ gauge field. All other monopole fluxes, in particular, all non-Abelian fluxes from the
SU(3) subgroup, are absorbed and squeezed in the confining  strings $S_1$, $S_2$ and $S_3$. 

\begin{figure}
\epsfxsize=10cm
\centerline{\epsfbox{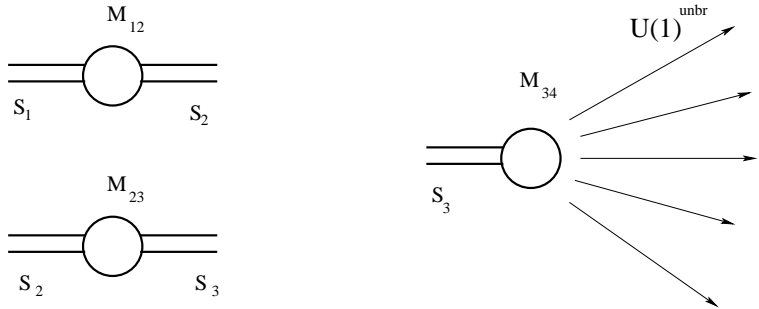}}
\caption{\small The monopole confinement in the $r=3$ vacuum. The thick double lines denote strings, while
the circles denote monopoles. Unconfined U(1) flux is shown by arrows.}
\label{figr=3conf}
\end{figure}

The picture of the monopole confinement in the $r=3$ vacuum is shown in Fig.~\ref{figr=3conf}.

To conclude this section let us determine the tensions of three elementary strings in the $r=3$ vacuum.
To the leading order in $\mu$, close to the \ntwo limit, these strings are BPS saturated. The Bogomol'nyi
representation for non-Abelian strings stabilized by the Fayet--Iliopoulos $F$-term is considered in 
\cite{SYfstr}. The boundary terms in this representation determine the string tensions,
\beq
T= {\rm Tr} \left\{ 
\left(
\begin{array}{ccc}
\xi_1 & \ldots & 0\\
\ldots &\ldots  & \ldots   \\
 0 & \ldots &  \xi_{N} \\
\end{array}
\right)
\,\int dx_i \left( \frac12 A_i
+T^a A_i^a\right) \right\}.
\label{bogten}
\eeq
The first diagonal matrix here is associated with quark condensates determined by $\xi$'s,
while the second matrix linear in $A$'s  represent 
the string flux. This formula is quite general and applies to any vacuum. Say, 
in the $r=N$ vacuum the  fluxes of the elementary $Z_N$ strings are \cite{SYrev,SYfstr}
\beq
\int dx_i \,{\rm diag}\left( \frac12 A_i
+T^a A_i^a\right)_{S_P} =
2\pi \,\left(0,...,1,0,...,0 \right)
\label{strfluxr=4}
\eeq
with the only nonvanishing element located at the $P$-th position, $P=1,...,N$. This implies the result  \cite{SYfstr}
for the tension of the $P$-th string quoted in (\ref{ten}).

In the $r=3$ vacuum at hand the string fluxes are determined by Eqs.~(\ref{S1}), (\ref{S23}).
Thus, we have 
\beqn
&&
\int dx_i \,{\rm diag}\left( \frac12 A_i
+T^a A_i^a\right)_{S_{1,2,3}} =
2\pi \,{\rm diag}\left\{ \left(1,0,0,\frac17 \right), \right.
\nonumber\\[2mm]
&&
\left.  \qquad\qquad \qquad\qquad\qquad \quad\left(0,1,0,\frac17 \right),
\quad \left(0,0,1,\frac17 \right)\right\} .
\label{strfluxr=3}
\eeqn
This gives the tensions for three elementary strings
\beq
T_{S_{1,2,3}}= 2\pi\,\xi_{1,2,3}.
\label{tenr=3}
\eeq
Note, that the last (nonvanishing) element  in (\ref{strfluxr=3}) (i.e. 1/7) 
does not contribute because of the condition $\xi_{N=4}=0$.

We see that the string tensions in the $r=N-1$ vacuum are still determined by nonvanishing $\xi$'s, 
in much the same way as in the
$r=N$ vacuum. In fact, we can fine-tune the quark masses in such a way  that the $r=3$ vacuum coalesces with 
the $r=4$ vacuum
(this amounts to taking $\xi_4\to 0$). Then Eqs.~(\ref{ten}) and (\ref{tenr=3}) show continuity of string tensions.

\section{Dual theory  in the \boldmath{$r=N-1$} vacuum}
\label{rdual}
\setcounter{equation}{0}

Now we will decrease the  parameter  $\xi$ passing in the domain of small 
$\xi$. Then the original theory (\ref{modelr}) finds itself in the  strong coupling regime. 
As we already explained in Sec.~\ref{rduality} ,(see also \cite{SYdual}) in order to study the 
transition from large to small
$\xi$ we first  assume the quark mass differences $\Delta m_{AB}=m_A-m_B$ to be large, 
$$| \Delta m_{AB}|\gg \Lambda_{{\mathcal N}=2}\,.$$ 
In this domain the theory stays at
weak coupling, and we can safely decrease the value of $\xi$. 

Next,
we  use the exact Seiberg--Witten solution of the theory on the Coulomb branch \cite{SW1,SW2}
 to pass  from the domain of the large quark mass differences to that with small quark mass differences, 
$$| \Delta m_{AB} |\ll \Lambda_{{\mathcal N}=2}\,.$$
In doing so we keep the quark masses themselves  large, 
$$|m_{A}| \gg \Lambda_{{\mathcal N}=2}\,.$$ 
In this limit
the non-Abelian subgroup of the low-energy  gauge group is U$(r=N-1)$ at large $\xi$ (see Sec.~\ref{r=N-1largexi})
and, therefore, the crossover to strong coupling as
well as duality in the $r$-vacuum look very similar to the those
in the  $r=N$ vacuum in the
U$(N)$ theory studied in \cite{SYdual}. 

Summarizing, 
 in this section we will assume the following conditions for the dual theory:
\beq
|\Delta m_{AB}| \ll \Lambda_{{\mathcal N}=2}, \qquad |m_{A}| \gg \Lambda_{{\mathcal N}=2}, \qquad
|\xi_P |\ll \Lambda_{{\mathcal N}=2}^2, \qquad |\mu |\ll \Lambda_{{\mathcal N}=2}.
\label{dualregion}
\eeq

To be more precise, the Seiberg--Witten curve factorizes in the $r=N-1$ vacuum in the folowing way \cite{CaInVa}:
\beqn
y^2&=& \prod_{k=1}^{N} (x-\phi_k)^2 -
4\left(\frac{\Lambda_{{\mathcal N}=2}}{\sqrt{2}}\right)^{2N-N_f}\, \,\,
\prod_{A=1}^{N_f} \left(x+\frac{m_A}{\sqrt{2}}\right)
\nonumber\\[4mm]
&=&\prod_{P=1}^{N-1} (x-e_P)^2\,(x-e_N^{+})(x-e_N^{-})\,.
\label{rcurve}
\eeqn
It has $r=(N-1)$ double roots associated with the quark condensation, so that
for the large mass differences $e_P$'s are given  by the
mass parameters, $\sqrt{2}e_P\approx -m_P$ ($P=1, ... , N-1)$. The last two roots (and $\phi_N$) are of order of 
$\Lambda_{{\mathcal N}=2}$. For single-trace deformation superpotential (\ref{msuperpotbr}), with
$\gamma=0$, (see (\ref{gamma})) their sum vanishes \cite{CaInVa},
\beq
e_N^{+} + e_N^{-}=0\,.
\label{DijVafa}
\eeq
This condition is equivalent to the following physical condition:
\beq
\xi_N= -2\sqrt{2}\mu\,E_N =0,
\label{xiN0}
\eeq
which is valid because the $N$-th quark is  heavy; therefore, it   develops no VEV. We already obtained this condition
in the classical limit, (see (\ref{xiN})). Below we will see that it is satisfied also in the
quantum theory.
The root $e_N^{+}$ determines the value of the gaugino condensate \cite{Cachazo2}.

Once $\Delta m_{AB}\ll \Lambda_{{\mathcal N}=2}$  (while 
$m_{A}\approx m \gg \Lambda_{{\mathcal N}=2}$)  $x$ is 
close to $-m/\sqrt{2}$, if we are interested in double roots of the curve. 
Then the curve can be approximately written as
\beqn
y^2&\approx &\left(\frac{m}{\sqrt{2}}\right)^2\,\left\{\prod_{k=1}^{r} (x-\phi_k)^2 -
4\left(\frac{\Lambda^{{\rm LE}}_{{\mathcal N}=2}}{\sqrt{2}}\right)^{2r-N_f}\, \,\,\prod_{A=1}^{N_f} \left(x+\frac{m_A}{\sqrt{2}}\right)\right\}
\nonumber\\[4mm]
&\approx &\left(\frac{m}{\sqrt{2}}\right)^2\,\prod_{P=1}^{r} (x-e_P)^2\, ,
\label{rcurveappr}
\eeqn
where the parameter $\Lambda^{{\rm LE}}_{{\mathcal N}=2}$ is given in (\ref{LambdaN2LE}).

We see that the curve reduces to the curve for the $r$-vacuum in the U$(r)$ theory. Now we  use the results obtained
in \cite{SYdual} where the transition to the strong coupling (small $\Delta m_{AB}$) was studied in this case.

To conclude this subsection we present, as an illustration,  the $\phi$ values  and roots of the 
curve (\ref{rcurve}) for the particular theory  (\ref{example}), in the limit of large  masses
(\ref{dualregion}). In this limit $\phi$'s are  
\beq
\phi_{1,2} = -\frac{m_{1,2}}{\sqrt{2}},\qquad \phi_3\approx -\frac1{\sqrt{2}}(m_3+
\Lambda^{{\rm LE}}_{{\mathcal N}=2}), \qquad \phi_4 \approx 0,
\label{r3phi}
\eeq
while the roots have the form
\beq
e_{1,2} =-\frac{m_{1,2}}{\sqrt{2}},\quad e_3\approx -\frac1{\sqrt{2}}(m_3-
\Lambda^{{\rm LE}}_{{\mathcal N}=2}), \quad e_4^{\pm} \approx \pm 
\sqrt{2m_3\Lambda_{{\mathcal N}=2}^{{\rm LE}}}\,.
\label{r3roots}
\eeq
Here we assume for simplicity that $m_4=m_1$ and $m_5=m_2$, cf. \cite{SYdual}.
We see that $e_{1,2}$ are exactly given by the masses (see Sec.~\ref{rduality}), while $e_4^{+}$ is much smaller than the double roots. 

\subsection{Monodromies}
\label{monodromies}

In this section we will study how quantum numbers of the massless quarks $q^{11}, ... ,\, q^{rr}$
in the $(1, ... ,\, r)$ vacuum
change as we reduce $\Delta m_{AB}$ to pass from weak coupling  to the
strong coupling domain along the Coulomb branch at $\xi=0$.

To simplify our  discussion we will consider 
a particular case (\ref{example}) 
so that the dual group has the smallest nontrivial rank $\nu=2$. We 
will consider the $(1, 2, 3)$ vacuum. The monodromies upon reducing the quark mass differences for the
$(1, 2, 3)$ vacuum
in the U(3) theory was studied in \cite{SYdual}. As was explained above, we can use these results for our $r=3$ vacuum in 
the U(4) theory if we keep $m \gg \Lambda_{{\mathcal N}=2}$.

The quark quantum numbers change  due to monodromies 
with respect to $\Delta m_{AB}\,$. The 
complex planes of $\Delta m_{AB}$ have cuts, and when we cross these cuts, the $a$ and $a_D$
fields acquire monodromies; the quantum numbers of the corresponding states change accordingly.
The method used in \cite{SYdual} to calculate the quark monodromies was the study of the Seiberg--Witten curve 
of the theory in the proximity of the
Argyres--Douglas  points \cite{AD} in $\Delta m_{AB}$ variables. In these AD points
 our 
$(1,2,3)$ vacuum collides with the monopole singularities. There are two relevant 
AD points for the theory at hand
\cite{SYdual}. The first one occurs at  
\beq
\Delta m_{31}=\Lambda_{{\mathcal N}=2}^{{\rm LE}}, \qquad 
e_1=e_3=-\frac{m_1}{\sqrt{2}}\,,
\label{AD1}
\eeq
where two double roots of the Seiberg--Witten curve (\ref{rcurveappr}) coincide, while the second is at
\beq
\Delta m_{32}=\Lambda_{{\mathcal N}=2}^{{\rm LE}}, \qquad e_2=e_3=-\frac{m_2}{\sqrt{2}}\,,
\label{AD2}
\eeq
where the other two double roots coincide. In these AD points 
the  monopoles $M_{13}$ and $M_{23}, respectively, $ become massless.
In \cite{SYdual} it was shown that passing through these AD points 
the quarks pick up magnetic charges of 
the corresponding monopoles, while 
the monopoles do not change their charges. As a result, below the AD points  the charges of the massless dyons are
\beqn
&& \vec{n}_{D^{1}} =
 \left(\frac12,0;\,\frac12,\frac12;\,\frac1{2\sqrt{3}},\frac{\sqrt{3}}{2};\,\frac1{2\sqrt{6}},0\right),
\nonumber\\[2mm]
&& \vec{n}_{D^{2}}=
\left(\frac12,0;\,-\frac12,-\frac12;\,\frac1{2\sqrt{3}},\frac{\sqrt{3}}{2};\,\frac1{2\sqrt{6}},0\right),
\nonumber\\[2mm]
&& \vec{n}_{D^{3}}=
\left(\frac12,0;\,0,0;\,-\frac1{\sqrt{3}},-\sqrt{3};\,\frac1{2\sqrt{6}},0\right),
\label{dyoncharges}
\eeqn
see (\ref{quarkcharges}) and (\ref{SU4roots}). 
Here we adjust results of \cite{SYdual} taking into account
the presence of the extra charge along $T^{15}$ in the U(4) theory. This amounts to just adding 
the quark charges with respect to this Cartan generator in (\ref{dyoncharges}),
 since  the $M_{13}$ and $M_{23}$ monopoles have no
$n_m^{15}$ charges, see (\ref{quarkcharges}) and (\ref{SU4roots}). 

Note, that as we decrease $\Delta m_{AB}$ we do not encounter other AD points in which the $M_{P4}$ monopoles 
 ($P=1, 2, 3$) become massless. To approach  these points one has to reduce $m_A$ (see (\ref{r3roots})),
 but we  keep $|m_A |$ large at the moment. 

Two remarks are in order here. First, it is crucially important
to note that the massless dyons 
$D^{1}$ and $D^{2}$ have both electric and magnetic charges $ 1/2$
with respect to the $T^3$ generator of the dual U$(\nu=2)$ gauge group. This means that they
can fill the fundamental representation of this group. Moreover, all dyons 
$D^{lA}$ ($l=1, ... ,\,\nu=2$) can form color doublets.
This is another confirmation 
of the conclusion made in Sect.~\ref{rduality}, that
the non-Abelian factor SU$(\nu=2)$ of the dual  gauge group gets restored 
in the equal mass limit.

A general  reason ensuring that the $D^{lA}$ ($l=1, ... ,\,\nu$) dyons  fill the fundamental
representation of the U$(\nu)$ group is as follows:
due to monodromies the $D^{lA}$ dyons  
pick up magnetic charges of particular  monopoles of SU($r$). 
The magnetic charges of these particular monopoles
are represented by weights rather than roots of the U$(\nu)$ subgroup ($\pm 1/2$ for U$(\nu=2)$,
see (\ref{SU4roots})). This is related to the absence of the
AD points associated with collisions of the first $\nu$ double roots, 
see (\ref{firstroots}). In other words, the dual 
SU$(\nu)$ theory is infrared-free and no monopole singularities occur in this subsector.
 
The second comment is that the
 dyon charges with respect to each U(1) generator are proportional to each other. This
 guarantees that these dyons are mutually local. 

\subsection{``Vacuum leap''}
\label{vjump}

In this section we will present the low-energy dual theory for the $r=N-1$ vacuum at small $\xi$. 
The gauge group of the theory is indicated in (\ref{rdualgaugegroup}).
 One of the U(1) factors of this group
corresponds to the unbroken U(1)$^{{\rm unbr}}$. The light matter sector consists of dyons which carry
weight-like electric charges as well as root-like magnetic charges. Non-Abelian dyons $D^{lA}$
($l=1, ... ,\nu$, $A=1, ... , N_f$) are in the fundamental representation of 
the SU$(\nu)$ dual gauge group.
There are also  dyon singlets  $D^{J}$ ($J=(\nu+1), ... , r$) charged with respect to 
the U(1) factors of the dual gauge group.
In the particular example 
(\ref{example}), the dyon charges were calculated in Sec.~\ref{monodromies}. In this example
we have a doublet of 
the non-Abelian dyons $D^{lA}$
($l=1, 2$) plus one singlet dyon $D^3$.
The  action of the dual theory for this case is presented in Appendix C.

The potential of this theory determines the dyons VEV's. In the generic $r=N-1$ vacuum we have
\beqn
\langle D^{lA}\rangle \! \! &=& \langle \bar{\tilde{D}}^{lA}\rangle =
\!\!
\frac1{\sqrt{2}}\,\left(
\begin{array}{cccccc}
0 & \ldots & 0 & \sqrt{\xi_{1}} & \ldots & 0\\
\ldots & \ldots & \ldots  & \ldots & \ldots & \ldots\\
0 & \ldots & 0 & 0 & \ldots & \sqrt{\xi_{\nu}}\\
\end{array}
\right),
\nonumber\\[5mm]
\langle D^{J}\rangle &=& \langle\bar{\tilde{D}}^{J}\rangle=\sqrt{\frac{\xi_J}{2}}, 
\qquad J=(\nu +1), ... , r\,.
\label{Dvevr}
\eeqn
In much the same way as in the $r=N$ vacuum
the most important feature  in (\ref{Dvev})  is a ``vacuum leap'' \cite{SYdual},
\beq
(1, ... ,\, r)_{\sqrt{\xi}\gg \Lambda_{{\mathcal N}=2}} \to (r+1, ... , \,N_f,\,\,(\nu+1), ... ,\, r)_{\sqrt{\xi}\ll \Lambda_{{\mathcal N}=2}}\,.
\label{jumpr}
\eeq
In other words, if we pick up the vacuum with nonvanishing VEVs of the  first $r$ quark flavors
in the original theory at large $\xi$,  and then reduce $\xi$ below 
$\Lambda_{{\mathcal N}=2}$, 
the system will go through a crossover transition and end up in the vacuum of the  dual theory with
the nonvanishing VEVs of $\nu $ last dyons (plus VEVs of $(r-\nu)$ SU$ (\nu)$ singlets).

\vspace{1mm}

The occurrence of this ``vacuum leap'' was demonstrated previously in \cite{SYdual} in  a particular example
of the $r=3$ vacuum in the U(3) gauge theory with $N_f=5$ flavors.
This was done  as follows. The curve (\ref{rcurveappr})
was studied with small mass differences $\Delta m_{14}$ and $\Delta m_{25}$. It was shown that
if at large $(m_3-m_{1,2})$ $\phi_{1,2}$ and $e_{1,2}$ were approximately given by $-m_{1,2}/\sqrt{2}$,
 respectively, then
at small $(m_3-m_{1,2})$ they approach $-m_{4,5}/\sqrt{2}$.

The $\xi_P$ parameters in (\ref{Dvevr}) can be calculated from the potential (\ref{potd}), see also (\ref{dbtoda}). It turns out that they are still determined by Eq.~(\ref{qxis}), 
in much  the same way as in the $r=N$ vacuum, 
where the matrix $E$ is given by (\ref{E}). However, for the $r=N-1$ vacuum the relation
(\ref{ErN})
between $E_P$ and the roots of the Seiberg--Witten curve modifies. 
In Appendix D we consider the simplest example of the $r=1$ vacuum in the
U(2) gauge theory to find this relation. An obvious generalization 
of the result (\ref{EeU2}) is
\beq
E_P=\sqrt{(e_P-e_N^{+})(e_P-e_N^{-})}, \quad P=1,...,(N-1), \quad E_N=0\,,
\label{Ee}
\eeq
which leads to our final expressions for the dyon VEVs  in terms of the
roots of the Seiberg--Witten curve,
\beq
\xi_P=-2\sqrt{2}\,\mu\,\sqrt{(e_P-e_N^{+})(e_P-e_N^{-})}, \quad P=1,...,(N-1), \quad \xi_N=0\,.
\label{xirN-1}
\eeq

Note that, at small $\Delta m_{AB}$, in the domain (\ref{dualregion}), the  first
$\nu$  roots are determined by the masses of the 
 last $\nu$ quarks, 
\beq
\sqrt{2}e_I = -m_{I+r}, \qquad 
I=1, ... ,\nu \,
\label{nuroots}
\eeq
(up to small corrections of order of $\Delta m^2/\Lambda_{{\mathcal N}=2}$).
 This is because  the 
non-Abelian sector of the dual theory is infrared-free and is at weak coupling
in the domain (\ref{dualregion}). 
As long as we keep $\xi_P$  small  the  dual theory is at weak coupling. For large masses
(see (\ref{dualregion})) this amounts to making $\mu$ sufficiently small.

\subsection{``Instead-of-confinement'' mechanism in \\ the
\boldmath{$r=(N-1)$} vacuum}
\label{ICr}

The phenomenon of the ``vacuum leap'' ensures that we have ``instead-of-confinement'' mechanism for 
the quarks and gauge bosons in the $r=(N-1)$-vacuum, in much  the same way as in the $r=N$ vacuum.

Indeed,
consider the mass choice (\ref{masssplitr}).
Both, the gauge group and the global flavor SU($N_f$) group, are
broken in the vacuum. However, the color-flavor locked form of (\ref{Dvevr})
shows that the  unbroken global group of the dual
theory is 
\beq
 {\rm SU}(r)_F\times  {\rm SU}(\nu)_{C+F}\times {\rm U}(1)\,.
\label{c+fdr}
\eeq
The SU$(\nu)_{C+F}$ factor in (\ref{c+fdr}) is a global unbroken color-flavor rotation, which involves the
last $\nu$ flavors, while the SU$(r)_F$ factor stands for the flavor rotation of the 
first $r$ dyons.

 In the equal mass limit, or given the mass choice (\ref{masssplitr}),
the global unbroken symmetry (\ref{c+fdr}) of the dual theory at small
$\xi$ coincides with the global group (\ref{c+fr})  in the
 the original theory at large $\xi$.  
However, again this global symmetry is realized in two different ways in the dual pair at hand.
The quarks and  gauge bosons of the original theory at large $\xi$
come in the $(1,1)$, $(r^2-1,1)$, $(\bar{r},\nu)$, and $(r,\bar{\nu})$
representations (see (\ref{onepr}), (\ref{twopr})), while the dyons and U($\nu$) gauge 
bosons form 
\beq
(1,1), \qquad (1,\nu^2-1)
\label{adjdual}
\eeq
and 
\beq
 (r,\bar{\nu}), \qquad  (\bar{r},\nu)
\label{bifunddual}
\eeq 
representations of (\ref{c+fdr}). We see again that the
adjoint representations of the $(C+F)$
subgroup are different in two theories.
     
This means that quarks and gauge bosons
which form the  adjoint $(r^2-1)$ representation  
of SU($r$) at large $\xi$ and the dyons and gauge bosons which form the  adjoint $(\nu^2-1)$ representation  of SU($\nu$) at small $\xi$ are different states. What happens to quarks and gauge bosons
at small $\xi$?

In much the same way as in the $r=N$ vacuum,
the screened  quarks 
and gauge bosons in the $r=(N-1)$ vacuum from the large-$\xi$ domain
 decay in the monopole-antimonopole  pairs on the CMS.
As we will show in Sect.~(\ref{monconfrd}), at small nonvanishing $\xi$ the
monopoles and antimonopoles produced in the decay process of the adjoint $(r^2-1,1)$ states
 are confined. Therefore, the (screened) quarks or gauge bosons 
evolve into stringy mesons in the  strong coupling domain of small  $\xi$ 
-- the  monopole-antimonopole
pairs connected  by two strings, as shown in  Fig.~\ref{figmeson}. 
The difference with ``instead-of-confinement'' phase in the $r=N$ vacuum is that in the $r=(N-1)$ vacuum
the strings can be broken by $M_{PN}$-monopole-antimonopole pairs (see the next subsection); here $P=1, ... , r$. As a
result, dipole stringy states emitting unbroken U(1)$^{{\rm unbr}}$ magnetic gauge fields
are formed, see Fig.~\ref{figdipole}. Non-Abelian SU$(\nu)$ fluxes are confined in these stringy dipoles.

Note, that in the large mass limit 
(\ref{dualregion}) the $M_{PN}$ monopoles  are very heavy, with masses of order of 
$m/g_2^2$; therefore, stringy mesons in Fig.~\ref{figmeson} are almost stable.

\subsection{Strings and monopole confinement in the dual \\ theory}
\label{monconfrd}

Now we will use the light dyon charges (\ref{dyoncharges}) to obtain the
 fluxes of the $Z_4$ strings in  the dual theory and show that these strings still confine monopoles.
 
Consider the $\tilde{S}_1$ string
arising due to winding of the  $D^{14}$ dyon. At $r\to\infty$ we have 
\beqn
D^{14}(r\to\infty) &\sim&
\sqrt{\frac{\xi_1}{2}}\,e^{i\alpha},
\qquad D^{25}(r\to\infty) \sim  \sqrt{\frac{\xi_2}{2}}, 
\nonumber\\[3mm]
 D^{3}(r\to\infty)&\sim&\sqrt{\frac{\xi_3}{2}},
\label{Dwind}
\eeqn
see (\ref{Dvevr}).
Note again that the condition associated with the fourth dyon is absent in the $r=3$ vacuum.
Taking into account the dyon charges  in Eq.~(\ref{dyoncharges}) 
we obtain the behavior of the gauge potentials at infinity,
\beqn
&& \frac12 A_i +\frac12 A_i^3 +\frac12 A_i^{3D}+ \frac1{2\sqrt{3}} A_i^8 
+ \frac{\sqrt{3}}{2} A_i^{8D} +\frac1{2\sqrt{6}} A_i^{15}\sim \pt_i \alpha\,,
\nonumber\\[2mm]
&& \frac12 A_i -\frac12 A_i^3 -\frac12 A_i^{3D}+ \frac1{2\sqrt{3}} A_i^8 
+ \frac{\sqrt{3}}{2} A_i^{8D} +\frac1{2\sqrt{6}} A_i^{15}\sim 0\,,
\nonumber\\[2mm]
&& \frac12 A_i  - \frac1{\sqrt{3}} A_i^8 - \sqrt{3} A_i^{8D} +\frac1{2\sqrt{6}} A_i^{15}\sim 0\,,
\eeqn
which, in turn, implies
\beqn
&&  A_i  +\frac1{\sqrt{6}} A_i^{15}\sim \frac23 \,\pt_i \alpha\,,
\nonumber\\[2mm]
&& \frac12 A_i^3 +\frac12 A_i^{3D}  \sim \frac12 \,\pt_i \alpha\,,
\nonumber\\[2mm]
&&   \frac1{2\sqrt{3}} A_i^8 
+ \frac{\sqrt{3}}{2} A_i^{8D} \sim \frac1{6}\,\pt_i \alpha \,.
\label{gaugewindint}
\eeqn
The combinations orthogonal to those which appear in (\ref{gaugewindint})
are required to tend to zero at infinity, 
namely, $A_i^3 - A_i^{3D}\sim 0$, $A_i^{8D} -3A_i^{8} \sim 0$ and $ A_i^{15D}\sim 0$.
Also taking into account (\ref{ort=0}) which stays intact in the dual theory we get 
\beqn
&&  A_i  \sim \frac47 \,\pt_i \alpha \,,\qquad A_i^D  \sim 0\,,
\nonumber\\[2mm]
&& A_i^3  \sim \frac12 \,\pt_i \alpha\,, \qquad A_i^{3D}  \sim \frac12 \,\pt_i \alpha\,,
\nonumber\\[2mm]
&&   A_i^8 \sim \frac1{10\sqrt{3}}\,\pt_i \alpha \,, \qquad A_i^{8D} \sim \frac{\sqrt{3}}{10}\,\pt_i \alpha\,
\nonumber\\
&&  A_i^{15}  \sim \frac4{7\sqrt{6}} \,\pt_i \alpha \,,\qquad A_i^{15D}  \sim 0\,.
\label{gaugewindd}
\eeqn
These expressions determine the charges of the $\tilde{S}_1$ string,
\beq
\vec{n}_{\tilde{S}_1}=
\left(0,\,\frac27;\,-\frac14,\,\frac14;\,-\frac{\sqrt{3}}{20},\,\frac1{20\sqrt{3}};\,0,\,\frac2{7\sqrt{6}}
\right).
\label{tS1}
\eeq

Paralleling the above analysis
we determine the charges of the other two  $Z_4$ strings which are due to windings of the fields
$D^{25}$ and $D^{3}$, respectively. We get
\beqn
&&
\vec{n}_{\tilde{S}_2}=\left(0,\,\frac13;\,\frac14,\,-\frac14;\,-\frac{\sqrt{3}}{20},\,\frac1{20\sqrt{3}};\,0,\,\frac2{7\sqrt{6}}\right),
\nonumber\\[3mm]
&&
\vec{n}_{\tilde{S}_3}=\left(0,\,\frac13;\,0,\,0;\,\frac{\sqrt{3}}{10},\,-\frac1{10\sqrt{3}};\,0,\,
\frac2{7\sqrt{6}}\right).
\label{tS23}
\eeqn
Now we can check that each of three  monopoles from 
the SU(3) subgroup of SU(4) can be confined by two strings. For the $M_{12}$ and $M_{23}$
monopoles    we have 
\beqn
&&
\vec{n}_{M_{12}}= (\vec{n}_{\tilde{S}_1}-\vec{n}_{\tilde{S}_2} )+
\frac12\left(\vec{n}_{D^{14}}-\vec{n}_{D^{25}}\right),
\nonumber\\[3mm]
&&
\vec{n}_{M_{23}}= (\vec{n}_{\tilde{S}_1}-\vec{n}_{\tilde{S}_2} )+
\frac3{10}\left(\vec{n}_{D^{25}}-\vec{n}_{D^{3}}\right) -\frac1{10}\left(\vec{n}_{D^{14}}-\vec{n}_{D^{25}}\right),
\label{confd}
\eeqn
where $\vec{n}_{D^{14}}$, $\vec{n}_{D^{25}}$ and $\vec{n}_{D^{3}}$ are charges of 
the condensed dyons  given in (\ref{dyoncharges}).
Only a part of the monopole flux is confined inside  the strings. The remainder  of its flux is
screened by the condensate of the $D^{14}$, $D^{25}$ and $D^{3}$ dyons. 

We see that, although the quark charges change as we pass from the large-$\xi$ domain  to small-$\xi$, and 
the quarks turn into dyons, this does {\em not} happen with the monopoles. The
monopole states do not change their charges. They are confined
in both, strong and weak coupling domains, being represented by the 
junctions of two different elementary strings. In the strong coupling domain in the dual theory 
there is a peculiarity: not the entire monopole flux is carried by  two attached strings; 
a part of it is screened by the dyon condensate.

Consider now the $M_{P4}$ monopoles ($P=1, 2, 3$). In much  the same way as in the original theory 
(see Sec.~\ref{monconf}), their fluxes in the dual theory  are not completely confined in 
the $r=3$ vacuum.
Consider, say, the $M_{34}$ monopole (see (\ref{M34})) attached to the string $\tilde{S}_3$. In 
the $r=3$ vacuum the $\tilde{S}_4$ string is absent due to the 
fact that $\xi_4=0$, and the flux of the above configuration is unconfined. 

Let us calculate this unconfined flux. It is easy to check that
\beqn
&&
\vec{n}_{\rm unconf} = \vec{n}_{\tilde{S}_3}-\vec{n}_{M_{34}}+
\frac1{10}\,\left(2\vec{n}_{D^{3}}-\vec{n}_{D^{14}}-\vec{n}_{D^{25}}\right)\,
\nonumber\\
&&
= \frac{2\sqrt{6}}{7}\,
\left(0,\,\frac1{\sqrt{6}};\,0,\,0;\,0,\,0;\,0,\,-1\right).
\label{unconffluxd}
\eeqn
Here we add in the right-hand side a linear combination of 
the charges the of $D^{14}$, $D^{25}$ and $D^{3}$ dyons. This linear combination 
is screened by their condensates. 
In much the same way as in the original theory, we see that the $n^8_m$ charge is canceled and the resulting charge is,
in fact, a source for the U(1) gauge 
magnetic field exactly  corresponding  to the field of the unbroken U(1)$^{\rm unbr}$ gauge group, see (\ref{unconf}). 

Thus, the $\tilde{S}_3$
string  can terminate on the monopole $M_{34}$ producing a magnetic source of the unbroken 
U(1)$^{\rm unbr}$ gauge field. All other monopole fluxes are absorbed by confining  
the $\tilde{S}_1$, 
$\tilde{S}_2$ and $\tilde{S}_3$ strings. The picture of the monopole confinement in 
the $r=3$ vacuum of the  dual theory is qualitatively the same as that in the original theory, see Fig.~\ref{figr=3conf}. 
Basically, the only difference is the fact that now confined non-Abelian fluxes are associated with the dual gauge group SU$(\nu=2)$, rather than with the original  SU$(r=3)$ group.

Note, that at large quark masses (see (\ref{dualregion}))  the $M_{P4}$ monopole masses ($P=1, 2, 3$) are very large;
 therefore, the  $\tilde{S}_P$ strings  are almost stable in this limit.

Note also, that, in much  the same way as 
in the original theory, the  tensions of $\tilde{S}_P$ strings are still given by Eq.~(\ref{tenr=3}),
where the $\xi_P$ parameters are determined by (\ref{xirN-1}).

\section{\boldmath{$r$}-Duality at large $\mu$}
\label{rdualitylargemu}
\setcounter{equation}{0}

Now we are ready to increase $\mu$ and decouple the adjoint matter. Our theory (\ref{model})
will flow to \none SQCD.

\subsection{Moving to the Argyres--Douglas point}
\label{ADregime}

In order to keep our dual theory at weak coupling we need to keep the $\xi$
parameters  (at least $\nu$ of them) sufficiently small. At large $\mu$ 
this creates a problem. In the $r=N$ vacuum this problem was overcame
in \cite{SYN1dual} by assuming the quark masses to be small. The $\xi$ parameters 
in the $r=N$ vacuum are 
given by (\ref{xirN}), while the first $(N_f-N)$ roots of the Seiberg--Witten curve are determined by 
the quark masses, with no $\Lambda_{{\mathcal N}=2}$-corrections, see (\ref{roots}). This allows us to
increase $\mu$ thus decoupling the adjoint matter as well as the U(1) factors, while keeping the 
low-energy U$(N_f-N)$ gauge  theory at week coupling.

Inspecting Eq.~(\ref{xirN-1}) we immediately see that this strategy does not work in the $r=N-1$ vacuum.

Although the first $\nu$ roots of the Seiberg--Witten curve are determined by the 
quark masses (see (\ref{nuroots})), the last two undouble roots $e_N^{\pm}$ are of order of 
$\Lambda_{{\mathcal N}=2}$
at small masses. Therefore, at large $\mu$ the $\xi$ parameters become large at small masses, 
$$\sim \mu \Lambda_{{\mathcal N}=2}\,,$$ destroying the weak coupling condition.

Thus, in the
$r<N$ vacua we need a different, novel  strategy. Equation (\ref{xirN-1}) shows that 
if we keep the mass differences very small and force the average value of the $\nu$ double roots  (determined by 
the quark masses,
that are almost equal) to lie in the proximity of one of the roots $e_N^{\pm}$, we make $\nu$ parameters $\xi$ small.
Say, we fine-tune the quark masses to ensure the limit
\beq
e_P \to e_N^{+}, \quad \Delta m_{KK'} \ll \Lambda_{{\mathcal N}=2}, \quad P=1, ... ,\nu, \quad K,K'= (r+1),...,N_f.
\label{ADlimit}
\eeq
Note, that it is possible to place all $\nu$ double roots close to $e_N^{+}$ because 
it is the quark masses rather 
than $\Lambda_{{\mathcal N}=2}$ that determine the ``non-Abelian''  roots of 
the Seiberg--Witten curve and the VEVs of the non-Abelian dyons, see (\ref{nuroots}).  

This limit means moving to the AD points. To see that this is indeed  the case observe
 that masses of $\nu$ monopoles $M_{PN}$
($P=1, ... ,\nu$) on the Coulomb branch are determined by the differences $(e_P - e_N^{+})\to 0$, 
the corresponding $\beta$-cycles shrink. 

Thus, besides the light dyons $D^{lA}$ and $D^J$ which are always present in our
$r$ vacuum, we get extra light monopoles  that are mutually nonlocal with the 
dyons. If we were on the Coulomb branch 
(at $\xi_P=0$) this  would definitely mean moving to strong coupling. In fact, the running coupling constant of our dual theory
is determined by the light dyon loops. If the monopoles simultaneously become light, their loops  give
logarithmic contributions to the inverse coupling, making the overall coupling constant of order of unity.

However, at small but nonvanishing $\xi$ we are {\em not} on the Coulomb branch. 
In fact, the monopoles are confined.
In particular, $\nu$ monopoles $M_{PN}$ ($P=1, ... ,\nu$) in question form stringy dipole states shown
in Fig.~\ref{figdipole}. Although the masses of the $M_{PN}$ monopoles  become very small in the limit 
(\ref{ADlimit}), the mass of the stringy dipole state formed by one of these monopoles (and an antimonopole) 
is determined by the string tension and, therefore, is much larger. It is of 
order of $\sqrt{\xi_P}$. The masses of the $D^{lA}$ dyons are 
of order of $\tilde{g}\sqrt{\xi}$. Starting from weak coupling in the dual theory 
and calculating the renormalization of the coupling constant $\tilde{g}$ we see 
that the monopole-antimonopole states are heavier, and their loops are suppressed. 
In the theory  (\ref{Sdual}) the coupling constant renormalization is determined by the dyon loops.
This ensures that the renormalized coupling constant is small, provided that we keep $\xi$'s small enough.

In other words, away from the Coulomb branch (at $\mu\neq 0$) the dual theory has no nontrivial
conformal AD-regime, which appears on the Coulomb branch in the limit (\ref{ADlimit}) \cite{AD}. 
It stays infrared-free. Note, however, that
the effective two-dimensional sigma model on the non-Abelian string goes 
into a nontrivial conformal regime at the AD-point \cite{Tongad}. 
This is because condensates of the scalar fields tend to zero inside 
the string core, and on the string we are essentially back to the Coulomb branch of the four-dimensional 
bulk theory.

Let us stress, that this is the most important observation which allows us to extend our $r$-duality
from \ntwo SQCD to \none.

The fact that  the light matter VEVs tend  to zero in the AD point was first recognized in \cite{GVY} 
in the  Abelian case.

\subsection{Decoupling the U(1) factors}

Now we can continue following the same road  as in \cite{SYN1dual}, where the
large-$\mu$ limit was studied in the $r=N$ vacuum. First we will 
take the limit (\ref{ADlimit}) still keeping $\mu$ small.

The VEVs of the non-Abelian dyons $D^{lA}$ become much smaller than the  VEVs of the
Abelian dyons $D^J$, see
(\ref{Dvevr}), (\ref{xirN-1}), and (\ref{nuroots}). In particular, the VEVs of the  $D^J$ dyons are 
determined by the differences $(e_J-e_N^{+})$ for $J=(\nu+1), ... , r$ which are not small and stay of order of 
$\Lambda_{{\mathcal N}=2}$ in the limit (\ref{ADlimit}). 

As a result,  $(N-\nu-1)$ U(1) gauge fields of the dual gauge group (\ref{rdualgaugegroup}) as well as
the  $D^J$ dyons themselves acquire large masses,
$\sim \sqrt{\mu \Lambda_{{\mathcal N}=2}}$,
 and decouple. At low energies we are left with the
\beq
{\rm U}(\nu)\times {\rm U}(1)^{\rm unbr}
\label{dualgaugelargemu}
\eeq
gauge theory of non-Abelian $D^{lA}$ dyons  ($l=1, ... ,\nu$, $A=1, ... , N_f$). The gauge field corresponding to
U(1)$^{\rm unbr}$ does not interact with the dyons and remains massless. 
The VEVs of the non-Abelian dyons are given by
\beq
\langle D^{lA}\rangle \! \! = \langle \bar{\tilde{D}}^{lA}\rangle =
\!\!
\frac1{\sqrt{2}}\,\left(
\begin{array}{cccccc}
0 & \ldots & 0 & \sqrt{\xi_{1}} & \ldots & 0\\
\ldots & \ldots & \ldots  & \ldots & \ldots & \ldots\\
0 & \ldots & 0 & 0 & \ldots & \sqrt{\xi_{\nu}}\\
\end{array}
\right),
\label{Dvevrlargemu}
\eeq
see (\ref{Dvevr}), where the first $\nu$ parameters $\xi_P$ are small in the limit (\ref{ADlimit}).

The superpotential of this theory can be written as
\beqn
&&
{\mathcal W} = \sqrt{2}\,\sum_{A=1}^{N_f}
\left( \frac{1}{ 2}\,\tilde D_A  b_{U(1)}
D^A +  \tilde D_A  b^p\,T^p  D^A
\right.
\nonumber\\[2mm]
&&
\left.
+ m_A\,\tilde D_A D^A \right)
+ \mu\,u_2( b_{U(1)}, b^p,a^{\rm unbr}).
\label{superpotdb}
\eeqn
Here $b_{U(1)}$ is a chiral superfield, the \ntwo superpartner of $B^{U(1)}_{\mu}$, 
where $B^{U(1)}_{\mu}$ is a particular linear combination of the dual gauge fields   not interacting
with the $D^J$ dyons. We normalized $b_{U(1)}$ so that the
charges of the $D^{lA}$ dyons 
with respect to this field
are  $\frac{1}{2}$.
This amounts to redefining its coupling constant $\tilde{g}^2_{U(1)}$.

 Moreover, 
$b^p$ (with $p=1, ... , \nu^2-1$) is an SU($\nu$) adjoint chiral field, 
the \ntwo superpartner of the dual SU($\nu$) gauge field,
see (\ref{Sdual}). We also use the
standard normalization for the non-Abelian charges of $D^{lA}$ absorbing
$\sqrt{2}$ present in (\ref{nablaD}) in the definition of the gauge fields. Finally, $a^{\rm unbr}$ is
a superpartner of the gauge field of the U(1)$^{\rm unbr}$, see (\ref{unbroken}).

\subsection{Decoupling adjoint matter}
\label{dam}

Now we increase $\mu$ and  make it
\beq
|\mu| \gg |\sqrt{\xi_P}|, \qquad P=1, ... , \nu
\label{muL}
\eeq
decoupling adjoint matter.  In order to keep the dual theory at weak coupling we 
go to the AD limit (\ref{ADlimit}) and require
\beq
|\sqrt{\xi_P}|\ll \tilde{\Lambda}, \qquad P=1,...,\nu\, ,
\label{weakcouprd}
\eeq
where 
\beq
\tilde{\Lambda}^{r-2\nu}= \frac{\Lambda_{{\mathcal N}=2}^{r-\nu }}{\mu^{\nu}}\,.
\label{tildeLr}
\eeq
We also assume that the quark mass differences are very small, even smaller than $E_P$, namely,
\beq
\Delta m_{KK'} \ll E_P=\sqrt{(e_P^2-e_N^{2})},\quad P=1,...,\nu, \quad K,K'= (r+1), ... , N_f\,.
\label{DeltamE}
\eeq

Given the  superpotential (\ref{superpotdb}) 
we can explicitly integrate out the adjoint matter. First we find 
the adjoint scalar VEVs. Say, in the simplest example $\nu=2$ we have
\beq
b^3=-\frac1{\sqrt{2}}\,\left(m_{N_f-1}-m_{N_f}\right), \qquad 
b_{{\rm U}(1)}=-\frac1{\sqrt{2}}\,\left(m_{N_f-1}+m_{N_f}\right).
\label{bvev}
\eeq

Next we find $a^{\rm unbr}$ from Eq.~(\ref{superpotdb})  and expand
the  resulting function $u_2$  in 
  powers of $b^p$ and deviations of $b_{{\rm U}(1)}$ from its VEV in (\ref{bvev}),
\beqn
u_2( b_{{\rm U}(1)}, b^p) 
&= & 
 c_1 \, (b^p)^2
+ c_2 \, \Delta b_{{\rm U}(1)} + c_3 \,  (\Delta b_{{\rm U}(1)})^2
\nonumber\\[3mm]
 &+&
O\left(\frac{\mu_2\, (b^p)^4}{\Lambda^2_{{\mathcal N}=2}}\right) + 
O\left(\frac{\mu_2\, (\Delta b_{{\rm U}(1)})^3}{\Lambda_{{\mathcal N}=2}}\right),
\label{supexpand}
\eeqn
Since  $\Delta b_{{\rm U}(1)}$ and $b^p$ are of order of $E_P$ (the VEVs of $b^p$ are also small, of order 
of $\Delta m_{KK'}$, see (\ref{bvev})) we can neglect higher-order
terms in the expansion (\ref{supexpand}) and keep only linear and quadratic 
terms. Higher-order terms are suppressed by powers of $E_P/\Lambda_{{\mathcal N}=2}$.

Now, substituting (\ref{supexpand}) into (\ref{superpotdb}) and integrating over
$\Delta b_{{\rm U}(1)}$ and $b^p$ we get the superpotential which depends only on $D^{lA}$. 
Minimizing it and 
requiring  the
VEVs of $D^{lA}$ to be given by  (\ref{Dvevrlargemu}) (see also (\ref{xirN-1})) we fix the coefficients
 $c_1$ and $c_2$. Say, for $\nu=2$ we get
\beq
c_1= -\frac{1}{2\sqrt{2}}\,\frac{\hat{m}}{\hat{E}}\,,\qquad
c_2= 2\hat{E}\,,
\label{cs}
\eeq
where
\beq
\hat{m}=\frac1{\nu}\,\, \sum_{P=1}^{\nu} m_{r+P}, \qquad 
\hat{E}=\frac1{\nu}\,\, \sum_{P=1}^{\nu} E_{P}\,.
\label{m}
\eeq
Note that the constant $c_3$ cannot be fixed by this procedure.  In principle, $c_3$ can be  fixed by studying
the behavior of $u_2$ near the AD points.

After eliminating the adjoint matter the superpotential takes the form
\beqn
{\mathcal W} &=& \frac{\hat{E}}{\sqrt{2}\,\hat{m}\,\mu}\,
 (\tilde{D}_A D^B)(\tilde{D}_B D^A) +\left[(m_A-\hat{m})+\frac{(\sqrt{2}\,\hat{E})^2}{\hat{m}}\right]\,
(\tilde{D}_A D^A)
\nonumber\\[3mm]
&+& 
c\left[\frac1{2\mu}\,(\tilde{D}_A D^A)^2 + \sqrt{2}\nu\,\hat{E}\,(\tilde{D}_A D^A)\right].
\label{superpotrd}
\eeqn
This equation presents 
our final large-$\mu$ result for the superpotential of the theory dual to \none SQCD in the (1, ... , $r$) vacuum. 
The constant $c\sim 1$ remains undetermined; it is related to $c_3$ above.

One can check that minimization of this superpotential leads to correct values of the dyon VEVs, 
Eq.~(\ref{Dvevrlargemu}).
The theory with the superpotential (\ref{superpotrd}) 
possesses many other vacua in which different dyons (and different number of dyons) develop VEVs.
We consider only one particular vacuum here.  As was explained in
Sec.~\ref{vjump}, if we  choose the (1, ... , $r$) vacuum in the original theory 
above the  crossover,  then we end up in the $(0, ... ,0, r+1, ... ,N_f)$ vacuum in the dual 
theory below the crossover, see (\ref{Dvevrlargemu}). Vacua with the number of condensed $D$'s
less than the maximum possible one (equal $\nu$)
seen in (\ref{superpotrd}) are spurious.

\subsection{Perturbative mass spectrum}
\label{pms}

Now we briefly summarize the perturbative mass spectrum of our dual theory with superpotential (\ref{superpotrd})
given  the quark mass choice (\ref{masssplitr}).

The U$(\nu)$ gauge group is completely Higgsed, and the masses of the gauge bosons 
are 
\beq
m_{{\rm SU}(\nu)}=\tilde{g}_2\sqrt{\xi}
\label{Wmassd}
\eeq
for the SU$(\tN)$ gauge bosons, and 
 \beq
m_{{\rm U}(1)}=\tilde{g}_1\, \sqrt{\frac{\nu}{2}}\,\sqrt{\xi}\,.
\label{phmassd}
\eeq
for the U(1) gauge boson. Here $\tilde{g}_1$ and $\tilde{g}_2$ are dual gauge couplings
for the U(1) and SU$(\nu)$ gauge bosons, respectively, while $\xi$ is a common value of  the first
$\nu$ parameters $\xi_P$ (see Eqs.~(\ref{xirN-1}) and (\ref{nuroots})),
\beq
\xi = -2\,\mu\,\sqrt{\hat{m}^2-2e_N^2}\,.
\label{xi}
\eeq
The dyon masses are determined by the $D$-term potential
\beq
 V^{\rm dual}_D =
 \frac{\tilde{g}^2_2}{2}
\left( \bar{D}_A T^p D_A -
\tilde{D}_A T^p \bar{\tilde{D}}^A \right)^2
+ \frac{\tilde{g}^2_1}{8}
\left(|D^A|^2 -|\tilde{D}_A|^2 
\right)^2\,
\label{Dtermpot}
\eeq
and the $F$-term potential following from the superpotential (\ref{superpotrd}). Diagonalizing the quadratic
form given by these two potentials we find that, out of $4\nu N_F$ real degrees of freedom 
of the scalar dyons, $\nu^2$  are eaten up in  the Higgs mechanism, $\nu^2-1$ real scalar dyons have 
the same mass as the non-Abelian gauge fields, Eq.
(\ref{Wmassd}), while one scalar dyon has the mass (\ref{phmassd}). These dyons are scalar superpartners 
of  the SU$(\nu)$  and U(1) gauge bosons in \none massive vector supermultiplets, respectively. 

Another $2(\nu^2-1)$
dyons form a $(1,\nu^2-1)$ representation of the global group (\ref{c+fdr}). Their  mass is as follows:
\beq
m_{(1,\nu^2-1)}=\frac{\hat{E}^2}{\hat{m}}\,,
\label{adjd}
\eeq 
while two real singlet dyons are heavier, their  mass 
 \beq
m_{(1,\,1)}\sim \hat{E}
\label{singld}
\eeq
is determined by the last term (the one with unknown coefficient) in (\ref{superpotrd}).
Here 
\beq
\hat{E}= \frac1{\sqrt{2}} \,\sqrt{\hat{m}^2-2e_N^2}\,,
\eeq
see (\ref{Ee}).

The masses of $4N\nu$ bifundamental fields  are given by the mass split of $r$ first and $\nu$ last
quark masses, see (\ref{masssplitr}),
\beq
m_{(\bar{r},\, \nu)}= \Delta m\,.
\label{bifundd}
\eeq 
All these dyons are the scalar components of the \none chiral multiplets.

We see that the masses of the gauge multiplets and those of chiral matter get a large split in the limit
of  large $\mu$ and small $\hat{E}$. Chiral matter becomes much lighter than
the gauge multiplets cf. \cite{SYnone,SYrev}.

\subsection{Summary}
\label{summary}

To summarize, at large $\mu$,  upon reducing $\xi$, the original \none SQCD in the $r=N-1$ 
vacuum undergoes  a crossover transition
at strong coupling. In the domain (\ref{weakcouprd}) in the vicinity of the
 AD points (\ref{ADregime}) it is described by the 
weakly coupled infrared-free  dual theory, U$(\nu)\times$U(1)$^{\rm unbr}$ SQCD, with   
$N_f$ light dyon flavors. 
 Condensation of the light dyons
$D^{lA}$ in this theory leads to formation of the non-Abelian strings and confinement of monopoles.
Quarks and gauge bosons of the original \none SQCD are in  the ``instead-of-confinement'' phase: they decay into the
 monopole-antimonopole pairs on CMS and 
form stringy mesons. In fact,  in the AD-regime (\ref{ADregime}) the $M_{PN}$ monopoles  ($P=1, ... ,\nu$)
become very light and, therefore, strings are unstable. As a result, stringy mesons shown in  Fig.~\ref{figmeson}
decay into stringy dipoles, see Fig~\ref{figdipole}. Stringy dipoles with non-trivial charges with respect to 
the SU$(r)$ part of the global group (for example from the adjoint representation) are stable.

\section{Conclusions}

Our main task was to extend non-Abelian duality, that was observed previously \cite{SYdual} in the $r=N$ vacuum,
to vacua with a smaller number of condensed quarks, which we referred to as the $r$ vacua. The second task was exploration of the confinement mechanism both in the original and dual theories, as it reveals itself in the $r$ vacua. As in \cite{SYdual}
we start from the \ntwo theory slightly deformed by the adjoint field mass parameter $\mu$ and study the transition from large values of the FI parameters $\xi$ to small values. At large $\xi$ it is the original theory that is weakly coupled. As we move to smaller $\xi$ the original theory becomes coupled exceedingly stronger. A dual description becomes more appropriate.
We identify the dual gauge group (which, surprisingly, is not the Seiberg dual group if $r<N$), dual matter and dual theory as a whole. Remarkably, the ``dual quarks" are {\em not} monopoles.  We identify an ``instead-of-confinement'' mechanism.

Then  we increase the deformation parameter $\mu$ 
and repeat the whole program. At large $\mu$ 
the adjoint fields decouple, and   our theory flows to \none SQCD. 
The gauge group of the dual theory becomes U$(N_f-r)$. We show that
the dual theory is still weakly coupled if we approach
the Argyres--Douglas point. The ``instead-of-confinement'' mechanism for quarks and gauge bosons 
survives in the limit of large $\mu$. It determines low-energy non-Abelian dynamics in 
the $r$-vacua of \none SQCD.

Our main example in this paper is the  $r=(N-1)$ vacuum. Still we expect that our results are quite general and 
can be applied to all $r > \frac23 N_f$ vacua. In particular, a generic $r$ vacuum has $(N-r-1)$ condensed monopoles at large $\xi$, in addition to $r$ condensed quarks. These monopoles are charged with respect to Abelian U(1) factors of the gauge group. At large $\mu$ and small $\xi$ in the dual theory all SU$(\nu)$ singlets
(including these monopoles) become heavy and decouple. They do   play no role in the low-energy dynamics
of the dual theory at large $\mu$. The light matter charged with respect to 
the dual gauge group U$(\nu)$ 
consists of the $D^{lA}$ dyons  which are quark-like states.  In particular,  
condensation of these dyons leads to confinement of monopoles.

A very crucial question is 
comparison of the $r$ duality we studied here with the Seiberg duality. This  will be carried out in a separate 
publication \cite{SYvsS}.

\section*{Acknowledgments}

 The work of MS was supported in part by DOE
grant DE-FG02-94ER40823. 
The work of AY was  supported
by  FTPI, University of Minnesota
and by Russian State Grant for
Scientific Schools RSGSS-65751.2010.2.

\section*{Appendix A:  \\
Low-energy action of the U\boldmath{$(N)$} theory in the \boldmath{$r=N-1$} vacuum at large 
\boldmath{$\xi$}}
\label{appB}
\addcontentsline{toc}{section}{Appendices}

 \renewcommand{\theequation}{A.\arabic{equation}}
\setcounter{equation}{0}
 
 \renewcommand{\thesubsection}{A.\arabic{subsection}}
\setcounter{subsection}{0}

 The low-energy action has the form
\beqn
S&=&\int d^4x \left[\frac1{4g^2_2}
\left(F^{n}_{\mu\nu}\right)^2 +
\frac1{4g^2_1}\left(F_{\mu\nu}\right)^2
+ \frac1{4g^2_2}\left(F^{(N^2-1)}_{\mu\nu}\right)^2
+\frac1{g^2_2}\left|D_{\mu}a^n\right|^2
\right.
\nonumber\\[4mm]
&+&
\left. \frac1{g^2_1}
\left|\partial_{\mu}a\right|^2 
+\frac1{g^2_2}
\left|\partial_{\mu}a^{(N^2-1)}\right|^2 
+ \left|\nabla_{\mu}
q^{A}\right|^2 + \left|\nabla_{\mu} \bar{\tilde{q}}^{A}\right|^2
+V\right]\,,
\label{modelr}
\eeqn
where the fundamental and  adjoint color  indices are $k=1, ... ,r$ and  $n=1, ... , r^2-1$,
respectively, while the
U(1) gauge field $A_{\mu}^{N^2-1}$ and its scalar superpartner $a^{N^2-1}$ are associated 
with the last Cartan generator of SU$(N)$. Note that
all non-Abelian gauge fields from the SU$(N)$/SU$(r)$ sector  are heavy and decouple in the large mass limit due to
 the structure of the adjoint VEVs, see (\ref{avevr}). Also the $q^{NA}$ quarks  are heavy and not included in 
 the low-energy theory.
The covariant derivative 
\beq
\nabla_\mu=\partial_\mu -\frac{i}{2}\; A_{\mu} -\frac{i}{\sqrt{2N(N-1)}}\; A_{\mu}^{N^2-1}
-i A^{n}_{\mu}\, T^n
\label{defnablar}
\eeq
acts in the fundamental representation.

\vspace{2mm}

The scalar potential $V(q^A,\tilde{q}_A,a^n,a,a^{(N^2-1)})$ in the action (\ref{modelr})
is
\beqn
&&
V(q^A,\tilde{q}_A,a^n,a,a^{(N^2-1)}) =
 \frac{g^2_2}{2}
\left( \frac{1}{g^2_2}\,  f^{nms} \bar a^m a^s
 +
 \bar{q}_A\,T^n q^A -
\tilde{q}_A T^n\,\bar{\tilde{q}}^A\right)^2
\nonumber\\[3mm]
&&
+\frac{g^2_1}{8}
\left(\bar{q}_A q^A - \tilde{q}_A \bar{\tilde{q}}^A \right)^2 
+\frac{g^2_2}{4N(N-1)}
\left(\bar{q}_A q^A - \tilde{q}_A \bar{\tilde{q}}^A \right)^2
\nonumber\\[3mm]
&&
+ 2g^2_2\left| \tilde{q}_A T^n q^A 
+\frac{1}{\sqrt{2}}\,\,\frac{\pt{\mathcal W}_{{\rm br}}}{\pt a^n}\right|^2+
\frac{g^2_1}{2}\left| \tilde{q}_A q^A +\sqrt{2}\,\,\frac{\pt{\mathcal W}_{{\rm br}}}{\pt a} \right|^2
\nonumber\\[3mm]
&&
+ 2g^2_2\left| \frac{1}{\sqrt{2N(N-1)}}\,\tilde{q}_A q^A 
+\frac{1}{\sqrt{2}}\,\,\frac{\pt{\mathcal W}_{{\rm br}}}{\pt a^{(N^2-1)}} \right|^2
\nonumber\\[3mm]
&&
+\frac12\sum_{A=1}^{N_f} \left\{ \left|\left(a+\sqrt{2}m_A +2T^n a^n+\sqrt{\frac{2}{N(N-1)}}a^{(N^2-1}\right)q^A
\right|^2\right.
\nonumber\\[3mm]
&&
+\left.
\left|\left(a+\sqrt{2}m_A +2T^n a^n+\sqrt{\frac{2}{N(N-1)}}a^{(N^2-1}\right)\bar{\tilde{q}}^A
\right|^2 \right\}\,.
\label{potr}
\eeqn

\section*{Appendix B:  \\[1mm]
Weights and roots of the SU(4) algebra}
\label{appA}

 \renewcommand{\theequation}{B.\arabic{equation}}
\setcounter{equation}{0}
 
 \renewcommand{\thesubsection}{B.\arabic{subsection}}
\setcounter{subsection}{0}

In this Appendix we present, for completeness,  weights and roots of 
the SU(4) algebra which we repeatedly use in the main text.
Weights determine quark charges, while roots determine monopole charges. The diagonal (Cartan) generators
of SU$(N)$ are defined as
\begin{eqnarray}
T^{\tilde{a}=(m+1)^2-1}_{ij}
&=&
\frac{1}{\sqrt{2m(m+1)}}\,\left(\sum_{k=1}^{m} \delta_{ik}\,\delta_{jk}
-m\,\delta_{i,m+1}\,\delta_{j,m+1}\right), 
\nonumber\\[2mm]
m
&=&
1, ... , N-1.
\end{eqnarray}
For SU(4) the index values $m=1,2,3$ corerespond to the Cartan generators $T^3$, $T^8$ and $T^{15}$.

In three-dimensional Cartan plane the weights of the SU(4) algebra are
\beqn
w_1 &=& \left(\frac12;\,\frac1{2\sqrt{3}};\,\frac1{2\sqrt{6}}\right) ,
\nonumber\\[2mm]
w_2 &=& \left(-\frac12;\,\frac1{2\sqrt{3}};\,\frac1{2\sqrt{6}}\right) ,
\nonumber\\[2mm]
w_3 &=& \left(0;\,-\frac1{\sqrt{3}};\,\frac1{2\sqrt{6}}\right) ,
\nonumber\\[2mm]
w_4 &=& \left(0;\,0;\,-\frac3{2\sqrt{6}}\right) .
\label{weights}
\eeqn
The
roots can be obtained as 
\beq
\alpha_{ij}=w_i-w_j, \qquad i<j.
\eeq
This implies
\beqn
\alpha_{12} &=& \left(1;\,0;\,0\right), \qquad 
\alpha_{13} = \left(\frac12;\,\frac{\sqrt{3}}{2};\,0\right) ,
\nonumber\\[2mm]
\alpha_{23} &=& \left(-\frac12;\,\frac{\sqrt{3}}{2};\,0\right), \qquad
\alpha_{14} = \left(\frac12;\,\frac1{2\sqrt{3}};\,\sqrt{\frac{2}{3}}\right) ,
\nonumber\\[2mm]
\alpha_{24} &=& \left(-\frac12;\,\frac1{2\sqrt{3}};\,\sqrt{\frac{2}{3}}\right), \qquad
\alpha_{34} = \left(0;\,-\frac1{\sqrt{3}};\,\sqrt{\frac{2}{3}}\right) .
\label{SU4roots}
\eeqn

For the monopole with charges determined by the root $\alpha_{ij}$ we use the notation $M_{ij}$.
From the expressions  above we find charges of all monopoles in SU(4). Say, for the  $M_{23}$ monopole  we 
have
\beq
\vec{n}_{M_{23}}=(0,\,0;\,0,\,-\frac12;\,0,\,\frac{\sqrt{3}}{2};\,0,\,0)
\eeq
in notations (\ref{chargenotation}).

\newpage

\section*{Appendix C:  \\[1mm]
Low-energy action of the dual theory  in the \boldmath{$r=3$} vacuum for \boldmath{$N=4$}}

 \renewcommand{\theequation}{C.\arabic{equation}}
\setcounter{equation}{0}
 
 \renewcommand{\thesubsection}{C.\arabic{subsection}}
\setcounter{subsection}{0}

The dual theory for the $r=3$ vacuum in the
U(3) gauge theory was found in \cite{SYdual}. To utilize these results in the $r=3$ vacuum in 
the U(4) theory  at hand we make a minor adjustment which takes into account the presence of an extra U(1) gauge field associated with  
the $T^{15}$ generator in the U(4) theory. The dual gauge group is U(2)$\times$U(1)$^{8}\times$U(1)$^{15}$. 
The bosonic part of the action is
\beqn
 S_{{\rm dual}} &=&\int d^4x \left[\frac{1}{4\tilde{g}^2_{2}}
\left(F^{p}_{\mu\nu}\right)^2 
+\frac1{4g^2_1}\left(F_{\mu\nu}\right)^2 +\frac1{4\tilde{g}^2_8}\left(F^8_{\mu\nu}\right)^2
+\frac1{4\tilde{g}^2_{15}}\left(F^{15}_{\mu\nu}\right)^2
\right.
\nonumber\\[4mm]
&+& 
\frac1{\tilde{g}^2_2}\left|\pt_{\mu}b^p\right|^2 
+\frac1{g^2_1}
\left|\partial_{\mu}a\right|^2 +\frac1{\tilde{g}^2_8}
\left|\partial_{\mu}b^8\right|^2
+\frac1{\tilde{g}^2_{15}}
\left|\partial_{\mu}a^{15}\right|^2
\nonumber\\[4mm]
&+&
\left.
\left|\nabla^1_{\mu}
D^A\right|^2 + \left|\nabla^1_{\mu} \tilde{D}_A\right|^2
\left|\nabla^2_{\mu}
D^3\right|^2 + \left|\nabla^2_{\mu} \tilde{D}_{3}\right|^2 + V\right]\,,
\label{Sdual}
\eeqn
Here covariant derivatives are defined in accordance
with the charges of the  $D^l$ ($l=1,2$) and $D^3$ dyons in (\ref{dyoncharges}). Namely,
\beqn
\nabla^1_\mu & = & 
=\pt_{\mu}-i\left(\frac12 A_{\mu}+\sqrt{2}\,B^p_{\mu}\frac{\tau^p}{2}
 +\frac{\sqrt{10}}{2\sqrt{3}}\,B^8_{\mu} +\frac1{2\sqrt{6}} A_{\mu}^{15}\right)\,,
\nonumber\\[3mm]
\nabla^2_\mu & = & 
=\pt_{\mu}-i\left(\frac12 A_{\mu}
 -\frac{\sqrt{10}}{\sqrt{3}}\,B^8_{\mu} +\frac1{2\sqrt{6}} A_{\mu}^{15}\right)\,,
\label{nablaD}
\eeqn
where the $B^{p}_{\mu}$ gauge fields  ($p=1,2,3$),  $B^{8}_{\mu}$, and their scalar superpartners $b^p$ and $b^8$
are 
\beqn
&&
B_{\mu}^3= \frac{1}{\sqrt{2}}\,(A_{\mu}^{3}+A^{3D}_{\mu}),\qquad 
b^3= \frac{1}{\sqrt{2}}\,(a^{3}+a^{3}_D)\;\;\; {\rm for}\;\;\; p=3,
\nonumber\\[4mm]
&&
B_{\mu}^8= \frac{1}{\sqrt{10}}\,(A_{\mu}^{8}+3A^{8D}_{\mu}),
\qquad b^8= \frac{1}{\sqrt{10}}\,(a^{8}+3a^{8}_D)\,.
\label{Bb}
\eeqn

The coupling constants $g_1$, $\tilde{g}_8$, $\tilde{g}_{15}$ and $\tilde{g}_2$ 
correspond to three U(1)'s and the SU(2) gauge groups, respectively.
The scalar potential $V(D,\tilde{D},b^p,b^8,a,a^{15})$ in the action (\ref{Sdual})
is 
\beqn
&& V =
 \frac{\tilde{g}^2_2}{4}
\left( \frac{1}{\tilde{g}^2_2}\,  f^{nms} \bar a^m a^s +\bar{D}_A\tau^p D_A -
\tilde{D}_A \tau^p \bar{\tilde{D}}^A \right)^2
\nonumber\\[3mm]
&+& \frac{10}{3}\frac{\tilde{g}^2_8}{8}
\left(|D^A|^2 -|\tilde{D}_A|^2 -2|D^3|^2 +
2|\tilde{D}_3|^2 \right)^2
\nonumber\\[3mm]
&+& \frac{\tilde{g}^2_1}{8}
\left(|D^A|^2 -|\tilde{D}_A|^2 +|D^3|^2 -
|\tilde{D}_3|^2 
\right)^2
\nonumber\\[3mm]
&+& \frac{\tilde{g}^2_{15}}{48}
\left(|D^A|^2 -|\tilde{D}_A|^2 +|D^3|^2 -
|\tilde{D}_3|^2 
\right)^2
\nonumber\\[3mm]
&+& \frac{\tilde{g}_2^2}{2}
\left| \sqrt{2}\tilde{D}_A \tau^p D_A +\sqrt{2}\,\,\frac{\pt{\mathcal W_{{\rm br}}}}{\pt b^p}
\right|^2+
\frac{\tilde{g}^2_1}{2}\left| \tilde{D}_A D^A+
\tilde{D}_3 D_3 +\sqrt{2}\,\,\frac{\pt{\mathcal W}_{{\rm br}}}{\pt a}\right|^2
\nonumber\\[3mm]
&+& 
\frac{\tilde{g}_8^2}{2}\left| \sqrt{\frac{10}{3}}\tilde{D}_A D^A-
2\sqrt{\frac{10}{3}}\tilde{D}_3 D^3 + \sqrt{2}\,\,\frac{\pt{\mathcal W}_{{\rm br}}}{\pt b^8}\right|^2
\nonumber\\[3mm]
&+&\frac{\tilde{g}^2_{15}}{2}\left| \frac1{\sqrt{6}}(\tilde{D}_A D^A+
\tilde{D}_3 D_3 ) +\sqrt{2}\,\,\frac{\pt{\mathcal W}_{{\rm br}}}{\pt a^{15}}\right|^2
\nonumber\\[3mm]
&+&\frac12 \left\{ \left|(a+\tau^p\sqrt{2}\,b^p +\sqrt{\frac{10}{3}}\,b^8+ \frac1{\sqrt{6}}\,a^{15}+\sqrt{2}m_A
)D^A\right|^2 
\right.
\nonumber\\[3mm]
&+& 
\left|(a+\tau^p\sqrt{2}\,b^p +\sqrt{\frac{10}{3}}\,b^8 + \frac1{\sqrt{6}}\,a^{15}+\sqrt{2}m_A
)\bar{\tilde{D}}_A\right|^2
\nonumber\\[3mm]
&+&\left.
\left|\;a-2\sqrt{\frac{10}{3}}\,b^8 + \frac1{\sqrt{6}}\,a^{15} +\sqrt{2}m_3 \;
\right|^2\left(|D^3|^2+|\tilde{D}_3|^2\right) \right\}\,,
\label{potd}
\eeqn
(see also \cite{SYN1dual}).

The derivatives of the superpotential ${\mathcal W}$ in (\ref{potd}) can be  calculated using
(\ref{dwda}). Next, we use monodromies found in the Sec.~\ref{monodromies} to relate 
the derivatives of
$u_2$ with respect to $b^3$ and $b^8$ to those with respect to $a^3$ and $a^8$, namely,
\beq
\frac1{\sqrt{2}}\,\frac{\pt u_2}{\pt b^3} = \frac{\pt u_2}{\pt a^3}, 
\qquad \frac1{\sqrt{10}}\,\frac{\pt u_2}{\pt b^8} = \frac{\pt u_2}{\pt a^8}\,,
\label{dbtoda}
\eeq
see also \cite{SYtorkink,SYN1dual}.

\section*{Appendix D:  \\[1mm]
The \boldmath{$r=1$} vacuum in U(2) theory}

 \renewcommand{\theequation}{D.\arabic{equation}}
\setcounter{equation}{0}
 
 \renewcommand{\thesubsection}{D.\arabic{subsection}}
\setcounter{subsection}{0}

In this Appendix we 
find the relation of the matrix $E$ (see (\ref{E}))    determining the quark/dyon VEVs in the original/dual theory with 
the roots of the Seiberg--Witten curve. We consider the simplest  possible example: 
the $r=1$ vacuum in the U(2) gauge theory with $1\le N_f<4$.

Let us calculate
the diagonal elements of the matrix $E$ given by
\beq
E=\frac1{2}\,\frac{\pt u_2}{\pt a}+\frac{\tau^3}{2}\,\frac{\pt u_2}{\pt a^3}
\label{EN2}
\eeq
in this particular case.
The Seiberg--Witten curve in this case factorizes as follows:
\beq
y^2=(x-e_1)^2\,(x-e_2^{+})(x-e_2^{-}),
\label{curveN=2}
\eeq
see (\ref{rcurve}). Here the double root at $x=e_1$ corresponds to a single condensed quark in 
the $r=1$
 vacuum, while
two other roots (subject to condition (\ref{DijVafa})) determine the gaugino condensate.

\begin{figure}
\epsfxsize=6cm
\centerline{\epsfbox{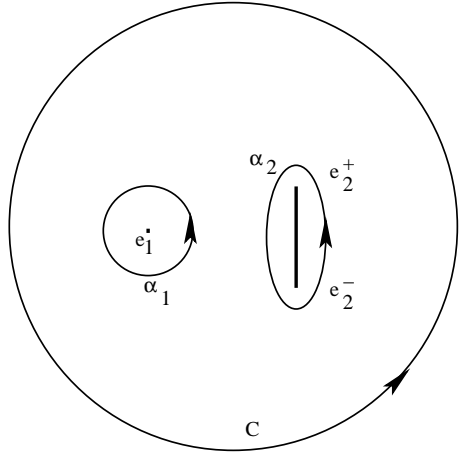}}
\caption{\small $\alpha$-contours in $x$-plane for the U(2) theory. Solid straight line denotes the cut. }
\label{figr=1contours}
\end{figure}

The exact solution of the theory on the Coulomb branch relates 
the fields $a$ and $a^3$ to 
contour integrals running along the contours $\alpha_i$ ($i=1,2$) in $x$-plane encircling  the double root $e_1$ and the cut which is stretched between the
roots  $e_2^{\pm}$, see Fig~\ref{figr=1contours}.

Using explicit expressions from \cite{ArFa,KLTY,ArPlSh,HaOz} and their generalization to 
the U$(N)$ case \cite{SYfstr}
we can write
\beq
\frac{\pt \Phi_i}{\pt u_2}= \frac12 \,\frac1{2\pi i}\oint_{\alpha_i}  \frac{dx}{y}\,,\,\,
\qquad \frac{\pt \Phi_i}{\pt u_1}=  \,\frac1{2\pi i}\oint_{\alpha_i} \frac{dx}{y}
\left[x-(e_1+e_2)\right]\,,
\label{dadu20}
\eeq
where the
variables $u_1$, $u_2$ are given by (\ref{u}), and we define
\beq
(\Phi_1,,...,\Phi_N)={\rm diag}\left(\frac12\, a + T^{\tilde{a}}\, a^{\tilde{a}}\right),
\label{Phii}
\eeq
while 
\beq
e_2=\frac12\left(e_2^{+}+e_2^{-}\right).
\label{eN}
\eeq
In fact, $e_2=0$ due to the condition (\ref{DijVafa}).

Equation~(\ref{Phii})  gives in the $N=2$ case 
\beq
a = \Phi_1+\Phi_2, \qquad a^3=\Phi_1-\Phi_2\,.
\label{PhiN2}
\eeq
For the factorized curve (\ref{curveN=2}) the integrals (\ref{dadu20}) can be  easily
evaluated. The integrals along the $\alpha_1$ contour are given by their pole contributions. To calculate 
the integrals along the $\alpha_2$ contour we write $\alpha_2= C -\alpha_1$, where $C$
is a large circle at infinity, see Fig~\ref{figr=1contours}.
This gives us
\beqn
&&
\frac{\pt \Phi_1}{\pt u_2} =\frac12\,\frac1{\sqrt{(e_1-e_2^{+})(e_1-e_2^{-})}},\qquad
\frac{\pt \Phi_2}{\pt u_2} =-\frac12\,\frac1{\sqrt{(e_1-e_2^{+})(e_1-e_2^{-})}},
\nonumber\\[3mm]
&&
\frac{\pt \Phi_1}{\pt u_1} =-\frac{e_2}{\sqrt{(e_1-e_2^{+})(e_1-e_2^{-})}},
\nonumber\\[3mm]
&&
\frac{\pt \Phi_2}{\pt u_2} =1+\frac{e_2}{\sqrt{(e_1-e_2^{+})(e_1-e_2^{-})}}\,.
\label{dPhidu}
\eeqn

 Using (\ref{PhiN2}) we get  the derivatives
$\pt a/\pt u_1$, $\pt a^3/\pt u_1$, $\pt a/\pt u_2$ and $\pt a^3/\pt u_2$.
Inverting this matrix and  substituting the result in   (\ref{EN2}) we  obtain
\beq
{\rm diag}\,E=\left(\sqrt{(e_1-e_2^{+})(e_1-e_2^{-})} +e_2,\,e_2\right)\,.
\eeq
Now we see that 
\beq
E_{N=2}=e_{N=2}=0,
\label{EN=eN}
\eeq
 i.e. the two conditions (\ref{DijVafa}) and (\ref{xiN0}) are equivalent.

Using these conditions we finally obtain
\beq
{\rm diag}\,E=\left(\sqrt{(e_1-e_2^{+})(e_1-e_2^{-})},\,0\right)\,.
\label{EeU2}
\eeq
Straightforward generalization of this result to arbitrary $N$ gives Eq. (\ref{Ee}) 
that was presented  in the main text.


\newpage 

\begin{thebibliography}{99}
\addcontentsline{toc}{section}{References}
\itemsep -2pt

\small

\bibitem{mandelstam}
Y.~Nambu,
  Phys.\ Rev.\  D {\bf 10}, 4262 (1974);\\
G.~'t Hooft,
{\em Gauge theories with unified weak, electromagnetic and strong interactions,}
in Proc. of the E.P.S. Int. Conf. on High Energy Physics, Palermo, 23-28 June, 1975
ed. A. Zichichi (Editrice Compositori, Bologna, 1976);
Nucl.\ Phys.\ B {\bf 190}, 455 (1981);
S.~Mandelstam,
Phys.\ Rept.\  {\bf 23}, 245 (1976).

\bibitem{SW1}
N.~Seiberg and E.~Witten,
Nucl. Phys. {\bf B426}, 19 (1994),
(E) {\bf B430},  485 (1994) [hep-th/9407087].

 \bibitem{SW2}
N.~Seiberg and E.~Witten,
Nucl. Phys. {\bf B431}, 484  (1994)
[hep-th/9408099].

\bibitem{DS}
M.~R.~Douglas and S.~H.~Shenker,
Nucl.\ Phys.\ B {\bf 447}, 271 (1995)
[hep-th/9503163].

\bibitem{HSZ}
A.~Hanany, M.~J.~Strassler and A.~Zaffaroni,
Nucl.\ Phys.\ B {\bf 513}, 87 (1998)
[hep-th/9707244].

\bibitem{Strassler}
M.~Strassler,
  Prog.\ Theor.\ Phys.\ Suppl.\  {\bf 131}, 439 (1998)
  [hep-lat/9803009].

\bibitem{VY}
A.~I.~Vainshtein and A.~Yung,
Nucl.\ Phys.\ B {\bf 614}, 3 (2001)
[hep-th/0012250].

\bibitem{SYdual}
  M.~Shifman and A.~Yung,
  Phys.\ Rev.\  D {\bf 79}, 125012 (2009)
[arXiv:0904.1035 [hep-th]].
  
\bibitem{SYN1dual}
  M.~Shifman and A.~Yung,
  Phys.\ Rev.\  D {\bf 83}, 105021 (2011)
  [arXiv:1103.3471 [hep-th]].

\bibitem{FI}
  P.~Fayet and J.~Iliopoulos,
  Phys.\ Lett.\  B {\bf 51}, 461 (1974).

\bibitem{HT1}
A.~Hanany and D.~Tong,
JHEP {\bf 0307}, 037 (2003)
[hep-th/0306150].

\bibitem{ABEKY}
R.~Auzzi, S.~Bolognesi, J.~Evslin, K.~Konishi and A.~Yung,
Nucl.\ Phys.\ B {\bf 673}, 187 (2003)
[hep-th/0307287].

 \bibitem{SYmon}
M.~Shifman and A.~Yung,
Phys.\ Rev.\ D {\bf 70}, 045004 (2004)
[hep-th/0403149].

\bibitem{HT2}
A.~Hanany and D.~Tong,
JHEP {\bf 0404}, 066 (2004)
[hep-th/0403158].
  
 \bibitem{Trev}
D.~Tong, {\em TASI Lectures on Solitons,}
  arXiv:hep-th/0509216.

\bibitem{Jrev}
  M.~Eto, Y.~Isozumi, M.~Nitta, K.~Ohashi and N.~Sakai,
  J.\ Phys.\ A  {\bf 39}, R315 (2006)
  [arXiv:hep-th/0602170].
  
  \bibitem{SYrev}
M.~Shifman and A.~Yung,
{\sl Supersymmetric Solitons,}
Rev.\ Mod.\ Phys. {\bf 79} 1139 (2007)
[arXiv:hep-th/0703267]; an expanded version in Cambridge University Press, 2009.

\bibitem{Trev2}
D.~Tong,
  Annals Phys.\  {\bf 324}, 30 (2009)
  [arXiv:0809.5060 [hep-th]].
  
  \bibitem{SYtorkink}
M.~Shifman and A.~Yung,
  Phys.\ Rev.\  D {\bf 81}, 085009 (2010)
  [arXiv:1002.0322 [hep-th]].

  \bibitem{SYcross}
M.~Shifman and A.~Yung,
Phys. Rev. {\bf D 79}, 105006 (2009)
  arXiv:0901.4144 [hep-th].

\bibitem{APS}
P.~Argyres, M.~Plesser and N.~Seiberg,
Nucl. Phys. {\bf B471}, 159  (1996)
[hep-th/9603042].

 \bibitem{Sdual}
  N.~Seiberg,
  Nucl.\ Phys.\  B {\bf 435}, 129 (1995)
  [arXiv:hep-th/9411149].
    
\bibitem{IS}
K.~A.~Intriligator and N.~Seiberg,
  Nucl.\ Phys.\ Proc.\ Suppl.\  {\bf 45BC}, 1 (1996)
  [hep-th/9509066].
  
  \bibitem{Cachazo2}
F.~Cachazo, N.~Seiberg and E.~Witten,
  JHEP {\bf 0304}, 018 (2003)
  [hep-th/0303207].

\bibitem{Ookouchi}
C.~Ahn, B.~Feng, Y.~Ookouchi, M.~Shigemori, 
Nucl. \ Phys. \ {\bf B698}, 3 (2004) 
[arXiv:hep-th/0405101]

\bibitem{BolKonMar}
S.~Bolognesi, K.~Konishi, G.~Marmorini,	
Nucl.  \ Phys. \ {\bf B718}, 134 (2005) 
[arXiv:hep-th/0502004].

\bibitem{AD}
P. C.~Argyres and M. R.~Douglas,
Nucl. \ Phys. \ {\bf B448}, 93 (1995)   
[arXiv:hep-th/9505062].
\\
P. C. Argyres, M. R. Plesser, N. Seiberg, and E. Witten,
Nucl. \ Phys.  \ {\bf B461}, 71 (1996) 
[arXiv:hep-th/9511154].


 \bibitem{SYvsS}
M.~Shifman and A.~Yung,
{\em Confronting Seiberg's duality with $r$ duality  in \none Supersymmetric 
QCD,}  arXiv:1204.4164[hep-th].

\bibitem{CKM}
G.~Carlino, K.~Konishi and H.~Murayama,
Nucl.\ Phys.\ B {\bf 590}, 37 (2000)
[hep-th/0005076].

\bibitem{ADS}
I.~Affleck, M.~Dine and N.~Seiberg,
  Nucl.\ Phys.\ B {\bf 241}, 493 (1984).
  
  \bibitem{Ven} 
  G.~Veneziano and S.~Yankielowicz,
  Phys.\ Lett.\ B {\bf 113}, 231 (1982).

\bibitem{GivKut}
A.~Giveon, D.~Kutasov,
Nucl. \ Phys.\  {\bf B796}, 25 (2008)
[arXiv:0710.0894[hep-th]].


\bibitem{SYfstr}
M.~Shifman and A.~Yung,
  Phys.\ Rev.\  D {\bf 82}, 066006 (2010)
  [arXiv:1005.5264 [hep-th]].

\bibitem{BF}
A.~Bilal and F.~Ferrari,
  Nucl.\ Phys.\  B {\bf 516}, 175 (1998)
  [arXiv:hep-th/9706145].

\bibitem{CaInVa}
F.~Cachazo, K.~A.~Intriligator and C.~Vafa,
  Nucl.\ Phys.\ B {\bf 603}, 3 (2001)
  [hep-th/0103067];
\\
V.~Balasubramanian, B.~Feng, M.~Huang and A.~Naqvi,
  Annals Phys.\  {\bf 310}, 375 (2004)
  [hep-th/0303065].

\bibitem{Tongad}
D.~Tong,
JHEP \ {\bf 0612}, 051 (2006) 
[hep-th/0610214]

\bibitem{ArFa}
P. C.~Argyres and A. E.~Faraggi,
Phys. \ Rev. \ Lett. {\bf 74},  3931 (1995)
[hep-th/9411057].

\bibitem{KLTY}
A.~Klemm, W.~Lerche, S.~Yankielowicz and S.~Theisen,
Phys. \ Lett. \ B {\bf 344}, 169 (1995) 
[hep-th/9411048].

\bibitem{ArPlSh}
P. C. Argyres, M. R. Plesser, and  A. Shapere,
Phys. \ Rev. \ Lett.  \ {\bf 75}, 1699 (1995)
[hep-th/9505100].

\bibitem{HaOz}
A.~Hanany and  Y.~Oz,
Nucl. \ Phys. \ B {\bf 452}, 283 (1995)
[hep-th/9505075].

\bibitem{GVY}
 A.~Gorsky, A.~I.~Vainshtein and A.~Yung,
  Nucl.\ Phys.\ B {\bf 584}, 197 (2000)
  [hep-th/0004087].

\bibitem{SYnone}
  M.~Shifman and A.~Yung,
  Phys.\ Rev.\ D {\bf 72}, 085017 (2005) [hep-th/0501211].



\end{thebibliography}
\end{document}